\documentclass[preprint,10pt,sort&compress]{elsarticle}

\usepackage{graphicx}
\usepackage{amssymb}
\usepackage{supertabular}
\usepackage{amsmath}
\usepackage[version=3]{mhchem}
\usepackage{bm}
\usepackage{braket}

\journal{Renewable Energy}

\begin{document}

\begin{frontmatter}



\title{Theoretical Limits of Photovoltaics Efficiency and Possible Improvements by Intuitive Approaches Learned from Photosynthesis and Quantum Coherence}

\author[add1,add2]{Fahhad H. Alharbi}
\ead{falharbi@qf.org.qa}
\author[add1,add3]{Sabre Kais}
\ead{sakias@qf.org.qa}
\address[add1]{Qatar Energy and Environment Research Institute (QEERI), Doha, Qatar}
\address[add2]{King Abdulaziz City for Science and Technology (KACST),Riyadh, Saudi Arabia}
\address[add3]{Department of Chemistry,  Physics and Birck Nanotechnology Center, Purdue University, West Lafayette, IN 47907 US}

\begin{abstract}

In this review, we present and discussed the main trends in photovoltaics with emphasize on the conversion efficiency limits. The theoretical limits of various photovoltaics device concepts are presented and analyzed using a flexible detailed balance model where more discussion emphasize is toward the losses. Also, few lessons from nature and other fields to improve the conversion efficiency in photovoltaics are presented and discussed as well. From photosynthesis, the perfect exciton transport in photosynthetic complexes can be utilized for PVs. Also, we present some lessons learned from other fields like recombination suppression by quantum coherence. For example, the coupling in photosynthetic reaction centers is used to suppress recombination in photocells.

\end{abstract}


\end{frontmatter}

\section{Introduction}

Sunlight is the most abundant energy source available on earth, and therefore to design systems that can effectively gather, transfer or store solar energy has been a great continuing interest for researchers. Maybe the most apparent field in this regard is photovoltaics (PV). Photovoltaics effect was known for about two centuries \cite{A01}. However, its serious technological development started in the 1950s. Various materials and device concepts have been developed since then and high conversion efficiencies have been achieved (44.7\% using quadruple junction \cite{F02}). This great development in the efficiency is not matched if the cost of the device is considered. The highly efficient PVs (mainly multi-junction solar cells) are prohibitively expensive \cite{J05,P04}. On the other hand, the efficiency of most dominant technology in the market (i.e. Si) is at 25\% in the lab and less than 20.\% commercially. In very interesting recent development, the hybrid perovskites solar cell (\ce{(CH3NH3)PbI3}) has attracted an extraordinary attention \cite{H02,J02,B02,L05,K06} as its efficiency has jumped to more than 16.2\% in about 4 years \cite{Added08}. Beyond that, the research trends have been wide spread though heavily material driven. One of the main research and development directions is to find cheaper and efficient absorbers. Other efforts focuses on developing alternative device concepts like multijunction and tandem solar cells. Reducing the ``fundamental'' losses is one of the main research field; but it proves to be very challenging.

Recently, new trends appeared to utilize intuitive approaches learned from other fields like photosynthesis and lasers.  In light harvesting organisms,  the major mechanism that converts light energy into chemical energy is photosynthesis. Remarkably, in plants, bacteria and algae, the photon-to-charge conversion efficiency is about $100\%$ under certain conditions \cite{Book1}. This fact is of great interest and generate a lot of excitement to understand how nature optimized the different molecular processes such as trapping, radiative, and non radiative losses, in particular the role of quantum coherence to enhance transport in photosynthesis \cite{added01,Added02,Added03}. This might lead to engineering new materials mimicking photosynthesis and could be used to achieve similar performances in artificial solar cells \cite{Book2,Creatore}. Quantum mechanics which was developed in the  twentieth century  continues to yield new fruit in the twenty-first century. For example, quantum coherence effects such as lasing without inversion \cite{Lasing,Lasing1}, the photo-Carnot quantum heat engine \cite{Carnot}, photosynthesis  and the quantum photocell \cite{Photocell} are topics of current research interest which are yielding new insights  into thermodynamics and optics.

In this review, we present collectively, different PV device concepts and the theoretical limits for their efficiencies where more discussion emphasize is toward the losses. In the analysis, a detailed balance model is used, where the balance is maintained between two-extended-level system that are affected by the solar radiation and the consequences like excitation and recombination \cite{H01,FH03}. The model is flexible and hence altered to accommodate all the analyzed device concepts. Then, we described  in some details photosynthesis and some quantum aspects from which lessons can be utilized in PV field.

\section{Photovoltaics: Alternative devices concept}
\label{PVDevConp}

\begin{figure} [t]
\centering
\includegraphics[width=2.6in]{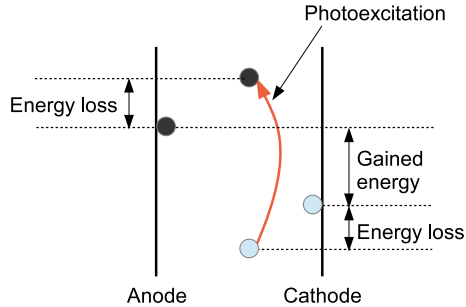}
\caption{The general concept of solar cell operation.}
\label{Concept}
\end{figure}

The general concept of solar cell is simple. An electron should be excited by solar radiation and then it should be collected at the anode before it losses the gained energy totally. Then the electron will be reinjected with energy below Fermi level $E_F$ into the cell from the cathode. The energy difference of the electron (between its energy at anode where it is collected and the energy at the cathode where it is reinjected) is used to do work (electrically, voltage times current). The cell should be designed so that the collection site (high-energy) can not supply carriers to the injection site (low-energy) as this will result in wasting the energy of the excited electron. The concept is shown schematically in Figure-\ref{Concept}.

\begin{figure} [t]
\centering
\includegraphics[width=3.0in]{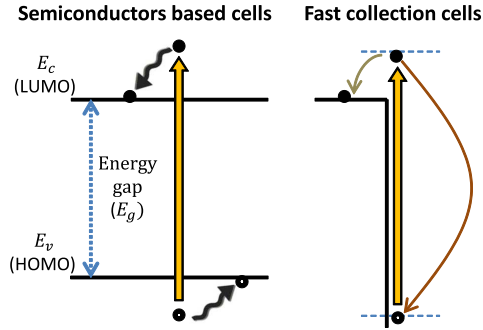}
\caption{The practical means to ensure gaining energy before it is lost totally. In the left, the gap in the semiconductors prevent the excited electron returning back to its originally lower energy position. In the right, the excited electron can be collected very rapidly by injection (green curved arrow) to electron carrier layer before it recombine (brown curved arrow).}
\label{SepConc}
\end{figure}

Conceptually, the semiconductors are not essential to realize photovoltaic effect though they are used in all solar cells now. In dye sensitized solar cells (DSSC), the semiconductors (i.e. \ce{ZnO} and \ce{TiO2}) are not used because of their semiconducting properties; they are merely used as an electron carrier and hole blocker. However, it is currently the most  convenient way to prevent losing all the energy gained by the excited electron. Practically there are two possible ways to ensure gaining energy; namely by the energy gap ($E_g$) in the semiconductors or very fast collection as shown in Figure-\ref{SepConc}. In the semiconductor, the excited carriers are relaxing back to the edge of the conduction and valance band. 

Photovoltaics effect was known for about two centuries. In 1839, Becquerel observed the effect accidentally while working on electrolytic cells \cite{B01,W01}. The first all-solid cell was made by Adams and Day in 1876 using selenium \cite{A01}. Later in the century, the first set of PVs patents appeared \cite{W02,W03,S01,S02}. Many efforts were conducted in the subject afterwards; but, the efficiency was extremely small. The practical realization was achieved in 1950, when a 6\% Si solar cell was made in Bell Labs \cite{C01} and then used for space applications. At that time, the work was based heavily on the conventional semiconductors and it was mainly to proof the concept. By 1960, 14\% efficient Si solar cells was made \cite{P01}; but, it was prohibitively expensive and not commercially lucrative. The need to reduce the cost lead to the second generation solar cells.

The work in the second generation solar cells started in the early 1960s where the aim was to reduce the fabrication cost of the solar cells in a trade of reduced efficiency. One of the most explored directions was to use alternative semiconductor absorbers. Tens of semiconductors were explored \cite{L01,W04,FH01,FH02,L02,C02} and the most prominent solar cells of that era were those based on \ce{CdTe} and \ce{Cu(In,Ga)Se2} (CIGS), which are thin film cells. Their current record efficiencies are 19.6\% for \ce{CdTe} \cite{G01} and 20.8\% for CIGS \cite{J01}. At the same era, Si based cell has been improved remarkably and its current efficiency record is 25\% which is about 8\% less than the theoretical limit \cite{S03,H01,FH03}. Some thorough theoretical analyses with more restricted practical assumptions indicated that the limit is not far above the obtained efficiency \cite{L03}.

Currently, we are in the midst of the third generation solar cell stage. The main aim of this stage is to make the electricity production cost of solar cell commercially competitive by reducing the cell fabrication costs and push the efficiencies above Shockley and Queisser limit \cite{W04,FH01,FH02,C02}. The developments have taken many directions, which can be categorized in different forms. In this paper, we categorize them based on the device concept as this is used later for the theoretical limits. In each category, some of the active research areas are briefly presented. Research activities, that are more toward conventional materials processing, are not addressed.

\subsection{Single junction devices}

\subsubsection{Alternative inorganic materials}
Conceptually, many inorganic semiconductors have the required physical properties to make solar cells \cite{L01,FH01,L02}. However, few of them have been extensively explored \cite{FH01,FH02,L02}. This area was very active in the 1970s and faded in the late 1980s. With the growing interest in solar energy, it has started gaining growing attentions in the recent years \cite{W04,FH01,FH02,C02}. The best obtained efficiencies of alternative absorbers are 17.1\% and 12.0\% for \ce{WSe2} \cite{P02} and \ce{MoSe2} \cite{P03}, respectively. For both of them, the device design was electrochemical cell. For all solid cell, the best efficiency is 8.0 for \ce{WSe2} absorber forming a heterojunction with \ce{ZnO} \cite{V01}.
	
\subsubsection{Organic photovoltaics (OPV)}
Organic semiconductors have been known for long time and they have been used in many relevant applications. For solar energy, Tang reported the first organic heterojunction solar cell in 1984 \cite{T01}. Since then, OPV field been very active especially in the past few years as the maximum obtained efficiency has been almost doubled \cite{D01,L04,C03,A02,S04} between 2009 (about 6\%) and now where the efficiency reached 11.1\% \cite{S04}.
	
\subsubsection{Sensitized solar cells} 
In 1988, Gratzel reintroduced the concept of dye sensitized solar cell (DSSC) with liquid electrolyte \cite{V02,G03,G04,O01}. It has attracted tremendous attention since then. the concept was introduced first by Gerischer \cite{G05,G06} and improved by Fujihira \cite{O02,F01}, Weller \cite{N01,K04}, and others. Practically, DSSC is a monolayer solar cell as the transport between the dyes is very small and having multiple dye layers causes a lot of practical challenges. Despite this fact, the device concept is very efficient and the latest obtained efficiency is 13.4\% \cite{N02}. Furthermore, enormous type of dye has been explored. One of the most interesting developments is the availability of solid based electrolyte \cite{C04,W05,A03,K04}. Also, it has been demonstrated that inorganic nanoparticles can replace the dye as sensitizer \cite{N03,G07,R01,K02,K03}. 
	
\subsubsection{Hybrid perovskites} 
Recently, the hybrid perovskites solar cell (\ce{(CH3NH3)PbI3}) has attracted an extraordinary attention \cite{H02,J02,B02,L05,K06} as its efficiency has jumped to more than 16.2\% in about 4 years \cite{Added08}. The absorbing perovskites are a special family of hybrid organic-inorganic crystalline materials with \ce{AMX3} perovskite structure, where \ce{A} is the organic site, \ce{M} is a metal, and \ce{X} is the halogen \cite{M02,M05,M01,B02}. The structure, which seems complex chemically, is extremely rich and can be grown and controlled relatively easily with high quality \cite{M02,M05,S05,C05,B03,L05}. It has been used for many applications \cite{M03,M04,K05} and its applicability for solar cells was anticipated very early \cite{M06}. However, the real thrust to make solar cells out of them is very recent \cite{M01,S05,H02,J02,B02,L05,K06}. In this short time, an efficiency of 16.2\% was reported \cite{Added08} and it is expected that 20\%+ efficiency can be achieved within few years \cite{M01,S05}.

\subsubsection{Nanostructured solar cells}
In the past two decades, the developments in nanotechnology has contributed a lot to introduce structures, materials, and mechanisms in solar cells, that are not possible in bulk form \cite{N03,L06,K02,K03,S06,A04,K07,H03,G08}. Among the effects are energy gap $E_g$ tunability, absorption and transport direction decoupling, and three-dimension structuring. However, due to the size related challenges, the obtained efficiencies are still small. Practically, nanostructuring results in deteriorated transport mainly due to interracial defects \cite{K08,L07,FH03, B04}. The best obtained efficiency in such structures is 9.0\% with \ce{PbS} quantum dots and \ce{TiO2} nano pillars by Sargent and coworkers [to be reported soon].

\subsection{Multi-cell devices}
	
\begin{figure*} [ht]
\centering
\includegraphics[width=6.0in]{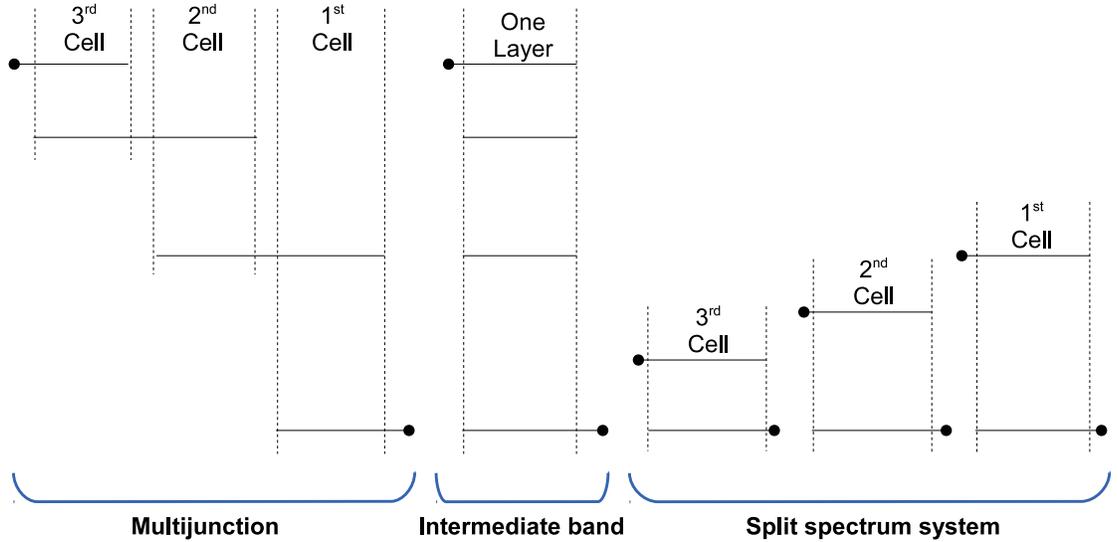}
\caption{The concepts of multijunction (left), intermediate band (center), and split spectrum (right) solar cell systems. The dots are the collection (injection) points.}
\label{MCSystems}
\end{figure*}
	
\subsubsection{Multijunction solar cells}
The long history and success of III-V optoelectronics allowed a smooth deployment of them in solar cell industry \cite{B05,Y01,Y02,D02} especially in the challenging structures like the multijunction cells. In such systems, few layers of different $E_g$ cells are stacked in series where in-between buffer layers allow transporting the photo-generated carriers between the layers. The system is two-terminal device as shown in Figure-\ref{MCSystems}. Its best obtained efficiency is 44.7\% \cite{F02} with quadruple junction developed by Fraunhofer ISE. However, there are many technological challenges that limit it such as the essentiality of current continuity, lattice matching, and the tunnelling of photo-generated carriers. Commercial wise, it is extremely expensive to fabricate \cite{J05,P04} and this fact limits it to limited applications. To distinguish it from the other concepts of multi-cell devices,  it is shown schematically represented in Figure-\ref{MCSystems} and as shown, it is conceptually a two-terminal system with a series stack of two-level cells. 
	
\subsubsection{Intermediate band cells}
	In such cells, the multi-photon absorption occurs in a single material layer and obviously, multi-level system is needed as represented in Figure-\ref{MCSystems}. The concept is relatively new and far from maturity as it is introduced my Luque and Marti in 1997 \cite{L06,L08,L09}. Although the concept is plausible, there are many challenges that should be resolved \cite{M07} and the obtained efficiencies based on this concept are small. Practically, there are two main trends to realize such system. The first one is based on doping large $E_g$ materials to create extended defect bands in the gap \cite{W06,L09,M08}. The second trend depends on super-lattices and organized quantum dots where many separated states can be created due to quantum size effects \cite{M09,L10,S07}.
	
\subsubsection{Split spectrum solar cell system}
The main idea of this system is to split solar radiation by a pre-optical setup and then direct each of the split spectrum into a cell with matching $E_g$ \cite{J03,M10,G09,FH04}. So, the system is composed of two parts. The first one is the optical system that splits the spectrum and concentrates the light. The second part is the set of SCs to be used to harvest the energy from the split spectrums as shown in Figure-\ref{MCSystems}. This avoids two of the main challenges that faces multijunction and intermediate band solar cells; namely current continuity and lattice matching in the case of multijunction cells and current continuity in intermediate band cells. The idea is not new and it has been suggested in 1955 \cite{J03} and patented in 1960 \cite{J04} by Jackson. Some some initial devices were developed in the 1970s. Moon et al demonstrated 28.5\% two-cell system in 1978 \cite{M10}. The record was set by Green and Ho-Baillie who obtained 43.5\% efficiency using 5-cell system \cite{G09}. Recently, it has been shown that a 50\%+ efficiency could be obtained using nowadays technologies \cite{FH04}. 

\subsection{Thermalization control based devices}

\subsubsection{Carrier multiplication devices} 
As known -and to be shown-, thermalization results in most losses in solar cells. The excess absorbed photon energy (above $E_g$) is lost as the hot carrier is relaxed into the band edge. The concept of carrier multiplication (CM) is based on utilizing  the energetic photon to generate multiple electron-hole pairs before it relaxes. Experimentally, CM has been demonstrated for both bulk and quantum sized semiconductor systems. Remarkably, Schaller and Klimov group achieved seven-fold multiplication in PbSe and PbS quantum dots (QDs) \cite{S08}. However, almost all CM experiments are done under unpractical conditions for solar cells. This fact and many other challenges have been highlighted repeatedly \cite{P05,B06,T02,FH03}.
	
\subsubsection{Hot carrier collection}
The idea of such cell concept is to collect the excited electron hot before it relaxes completely \cite{G10,L11}. In principle, this can be achievable by enabling very fast photo-current collection, using selected contacts, or slowing down the relaxation .\cite{W07,C06,C07}. Recently, it has been demonstrated that some nano materials like graphene enables -expectedly- hot carrier transport \cite{G11,S09,S10}.

\subsection{Spectrum manipulation based devices}

\subsubsection{Up- and down-conversion cells}
Solar radiation has a very broad spectrum from far infra-red into ultraviolet (around 4 eV). So, designing a device that utilize all the possible energy is challenging. The idea of up- and down- conversion cells is to manipulate the spectrum by various optical nonlinear systems to reduce the width of the resulted spectrum and then use the proper cell for energy harvesting \cite{A05,V03}. Such spectrum manipulation can extremely reduce the losses due to thermalization. However, the nonlinear conversion is a challenge by itself.  

\subsubsection{Thermophotovoltaics (TPV)}
The idea of such device is to utilize the generated heat (by photovoltaic losses) to generate extra electricity beside the photovoltaic output \cite{C08,C09,B07,D03,D05}. So, it consists basically of a thermal emitter and a photovoltaic. For the thermal emitter and to create more heat differential, it is common to use optical concentration with the system. The theoretical limit is far beyond that of the solar cells and many analyses show that the limit is just above 80\% \cite{D03,D04,B08,A06} (this is far beyond solar cell limits). The area is rich and many device designs and materials have been explored. However, the reported efficiencies are small \cite{D03,D04}.

\section{Energy conversion theoretical limits for various PV device concepts}
Theoretically, many models were used to estimate the maximum possible efficiencies of the solar cells. They can be categorized in two general families. The first category analyses are phenomenologically based on detailed balance of radiations between two-extended-level system. This accounts for excitation and radiative recombination. Originally, this was introduced by Shockley and Queisser in 1961 \cite{S03} and then followed by many others \cite{H01,FH03,Added04,Added05,Added06,Added07}. In this review, we use a model of this category \cite{H01,FH03}. It will be presented later. The second category is more fundamental and it is fully thermodynamical. These models are based on maintaining the balance of both energy and entropy fluxes \cite{L03,L12,L13}.

Practically and as mentioned in the previous subsection, there are many possible operation concepts of solar cells. To estimate the upper limit of each device concept, different assumptions are made. Here listed are the main assumptions that are taken in commonly considerations:
\begin{itemize}
\item Solar radiation strength and spectrum vary based on the position and system design. For example, the spectrum on satellites is different from that on earth surface as some spectrum lines are absorbed by gases on the atmosphere. Also, the strength and the spectrum can be altered by using pre-optical systems like concentrator and spectrum manipulation. This can be calculated from the base solar radiation. In this review, we assume AM1.5G photon flux ($\phi_{1.5}$) where the reference solar spectra ASTM G-173-03 (American Society for Testing and Materials) \cite{A07,R02} is used.
\item Any incident photon above the energy gap $E_g$ of the used cell is absorbed.
\item Any photon (with energy $E$) shall produce $\gamma(E)$ electrons, where $\gamma$ is the multiplication factor. In most cases, $\gamma=1$. Yet, if carrier multiplication is possible, it can take higher values. This will be included in the analysis.
\item There are many recombination mechanisms. Many of them -and unfortunately, the most effective ones- are caused by material quality, device design, and fabrications. Such non-fundamental mechanisms do not set the upper limit. The main unavoidable recombination mechanism is the one due to spontaneous emission. This is governed by the generalized black body radiation as will be shown later. In this work, this mechanism is forced as it is inevitable.
\end{itemize}

\subsection{Single junction solar cells}
In this review, a detailed balance model is used to estimate the upper efficiency limits under different conditions \cite{H01,FH03}. The balance is applied to the radiations in two-extended-level system. The first studied device structure is for single junction solar cells. As a radiation of flux $\phi (E)$ reaches the cell, the photo-generated current is then
\begin{equation}
\label{Jg}
	J_g (E_g) = q \int_{E_g}^\infty \gamma(E) \phi (E) dE
\end{equation}
where $E_g$ is the energy gap (in $eV$), $q$ is the electron charge, $E$ is photon energy (in $eV$), and $\gamma(E)$ is the multiplication as mentioned above. $\phi (E)$ in this case can be any manipulation of the base AM1.5G standard flux. Two cases will be considered in this work. The first one is to assume that the whole flux get to the cell without concentration and loss. The second case is with uniform optical concentration where the flux is simply multiplied by a factor $X$ that represent the uniform concentration.

As the recombination is assumed to be only due to spontaneous emission which is governed by the generalized black body radiation, the recombination current can be calculated accordingly and it is 
\begin{equation}
\label{Jr}
	J_r (E_g,V,T) = q a \int_{E_g}^\infty  \frac{E^2}{\exp\left( \frac{E - \gamma(E) V}{k T} \right) - 1}  dE
\end{equation}
where
\begin{equation}
	a = \left( \frac{2 \pi q^3}{c^2 h^3} \right),
\end{equation}
$c$ is the speed of light in vacuum in $m/s$, $h$ is Planck's constant in $eV \cdot s$, $V$ is the photo-generated voltage across the cell in $V$, $k$ is Boltzmann’s constant (in $eV/K$), and $T$ is the temperature (in $K$). So, the net current is then just the remaining photo-generated current after the recombination losses. So, 
\begin{equation}
\label{J}
	J \left( E_g, V, T \right) = J_g \left( E_g \right) - J_r \left( E_g, V, T \right)
\end{equation}
Then, the conversion efficiency can be calculated directly as the ratio between the output and input power. 
\begin{equation}
\label{Eff}
	\eta \left( V \right) = \frac{P_{out}}{P_{in}} = \frac{V \, J \left( E_g, V, T \right)}{P_{in}}
\end{equation}
where $P_{in}$ is the input power and it equals to
\begin{equation}
\label{Pin}
	P_{in} = q \int_0^\infty E \, \phi (E) dE
\end{equation}
The maximum possible efficiency at given $E_g$ and $T$ is obtained by varying $V$ to maximize $\eta$.

There are three main causes of losses in this model. The first one is due to the unabsorbed photons where its loss fraction is
\begin{equation}
	\label{LuaSJ}
	L_{unabs}^{(SJ)} = \frac{q}{P_{in}} \int_0^{E_g} E \, \phi (E) dE
\end{equation}
The second cause is due to thermalization where the loss is the difference between the energy of the absorbed photon (i.e. $E$) and the energy gained by its photo-generated electrons (i.e. $\gamma(E) V$). So,
\begin{equation}
	\label{LthSJ}
	L_{th}^{(SJ)} = \frac{q}{P_{in}} \int_{E_g}^\infty \left(E- \gamma(E) E_g \right) \, \phi (E) dE
\end{equation}
The last loss is to recombination and it is
\begin{equation}
	\label{LrSJ}
	L_{r}^{(SJ)} = \frac{V J_r}{P_{in}} + \frac{q}{P_{in}} \int_{E_g}^\infty \left(E_g - V \right) \gamma(E) \, \phi (E) dE
\end{equation}
The first term is the direct loss due to the recombination. The second term accounts for the further thermalization as a result of the balance between absorption and re-emission where the extracted photo-generated current gained $V$ potential instead of separation energy after initial relaxation (i.e. $E_g$).

\begin{figure} [ht]
\centering
\includegraphics[width=3.0in]{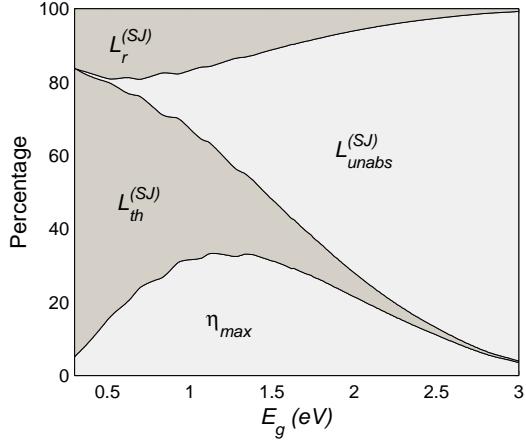}
\caption{The optimized $\eta$ of single junction solar cells for different $E_g$'s at $T = 300 K$ and with no carrier multiplication. The corresponding $L_{unabs}^{(SJ)}$, $L_{th}^{(SJ)}$, and $L_{r}^{(SJ)}$ for each $E_g$ are added up.}
\label{SJSCEffLosses}
\end{figure}

In the first analysis, $\eta$ is optimized for operation at room temperature ($T = 300 K$) for different $E_g$'s and with no carrier multiplication (i.e. $\gamma=1$). The corresponding  $L_{unabs}^{(SJ)}$, $L_{th}^{(SJ)}$, and $L_{r}^{(SJ)}$ are calculated as well as shown in Figure- \ref{SJSCEffLosses}. The maximum obtainable efficiency is 33.3\% at $E_g = 1.14 \; eV$, which is very close to the silicon $E_g$. However, there is a wide window of energy gaps between $0.91 \; eV$ and $1.57 \; eV$ that has efficiency limit about 30\%. Many semiconductors have gaps in this range and can conceptually be used to develop relatively efficient single-junction solar cells providing that the transport is good. 

Over the whole range of $E_g$, more that 50\% of energy is lost either due to thermalization or for not absorbed photons. This particular fact is the essence of many efforts to increase conversion efficiency. This is why device concepts such as multi-junction, split-spectrum, hot carrier, and carrier multiplication cells were introduced. Eq. \ref{LuaSJ} and Eq. \ref{LthSJ} show that these losses are -ideally- independent of temperature and the balance between absorption and re-emission (characterized by the gained potential $V$). Furthermore and ideally, they both are not affected by uniform optical concentration, where $\phi$ is simply concentrated to $X\phi$. So, all quantities depend linearly on $\phi$ and will be affected accordingly. From, Eq. \ref{Pin}, Eq. \ref{LuaSJ}, and Eq. \ref{LthSJ}, it can be shown that both $L_{unabs}^{(SJ)}$ and $L_{th}^{(SJ)}$ remain constant with uniform optical concentration. So, before considering the other device concepts, we will consider the effects of temperature and optical concentration on single junction cell efficiency limits and mainly on $\eta_{max}$ and $L_{r}^{(SJ)}$. 

\begin{figure} [ht]
\centering
\includegraphics[width=3.0in]{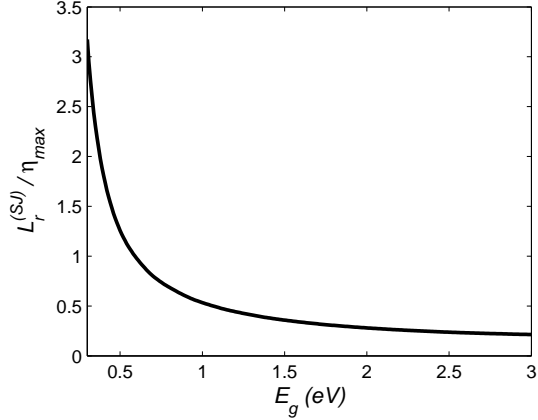}
\caption{The ratio between $L_{r}^{(SJ)}$ and $\eta_{max}$ of single junction solar cells for different $E_g$'s at $T = 300 K$ and with no carrier multiplication.}
\label{SJSCLTHoEta}
\end{figure}

First, it can be observed from Figure-\ref{SJSCEffLosses} that the recombination loss is quite large at small $E_g$ and it then gets reduced. Figure-\ref{SJSCLTHoEta} shows the ratio between $L_{r}^{(SJ)}$ and $\eta_{max}$. At small $E_g$ (up to about $0.59 \; eV$), the losses due to recombination are more than the obtained conversion. Even at higher $E_g$, a good portion of energy conversion is lost due to the recombination.

\begin{figure} [ht]
\centering
\includegraphics[width=3.0in]{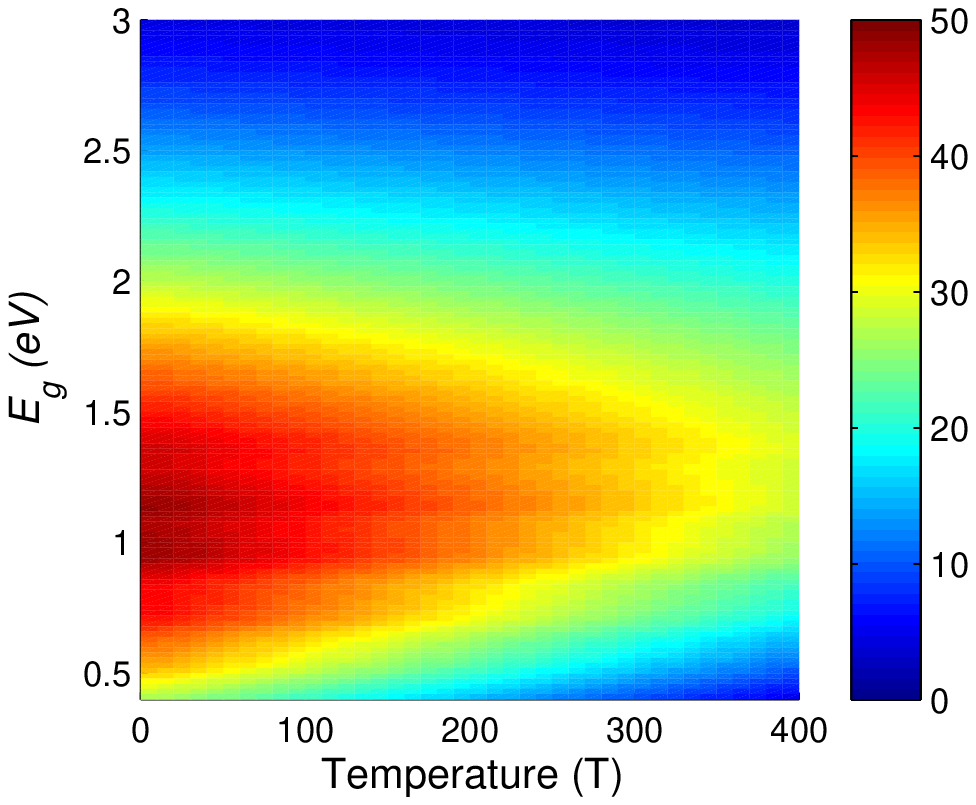}
\caption{The contour of $\eta_{max}$ vs. $E_g$ and temperature for a single junction solar cell and with no optical concentration.}
\label{SJSCTEff}
\end{figure}

\begin{figure} [ht]
\centering
\includegraphics[width=3.0in]{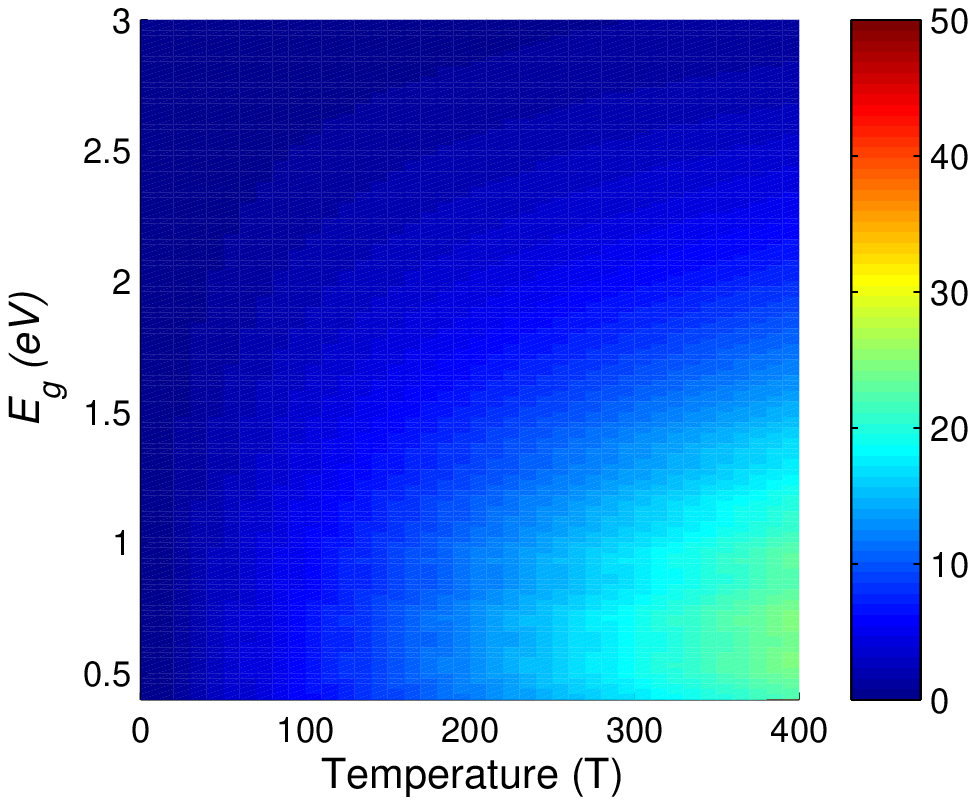}
\caption{The contour of $L_{r}^{(SJ)}$ vs. $E_g$ and temperature for a single junction solar cell and with no optical concentration.}
\label{SJSCTLTD}
\end{figure}

From Eq. \ref{LrSJ}, it is clear that the main cause for $L_{r}^{(SJ)}$ is the recombination current $J_r$, in which the denominator of the integrand depends on $T$ (Eq. \ref{Jr}). To absorb the photon, $E$ should be greater than $\gamma(E) V$. So, the integrand will increase with $T$ and hence $J_r$. So, $\eta_{max}$ should improve as $T$ decreases. This is actually the case as shown in Figure-\ref{SJSCTEff} as $L_{r}^{(SJ)}$ decreases with $T$ (Figure-\ref{SJSCTLTD}). At very low temperature, the maximum efficiency limit is 48.48\% at $1.12 \; eV$ gap. However, the window of $E_g$ of the highest obtainable conversion is slightly red-shifted to become between $0.86 \; eV$ and $1.40 \; eV$ for limits about 45\%.

\begin{figure} [ht]
\centering
\includegraphics[width=3.0in]{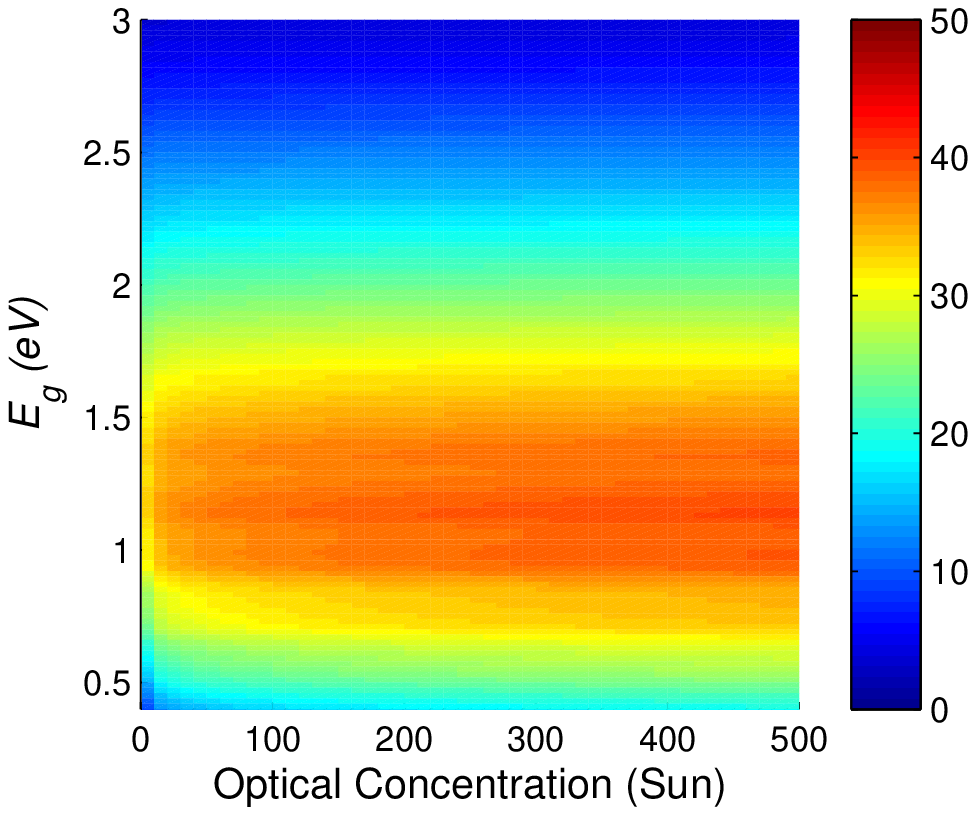}
\caption{The contour of $\eta_{max}$ vs. $E_g$ and $X$ for a single junction solar cell at 300 K.}
\label{SJSCXEff}
\end{figure}

\begin{figure} [ht]
\centering
\includegraphics[width=3.0in]{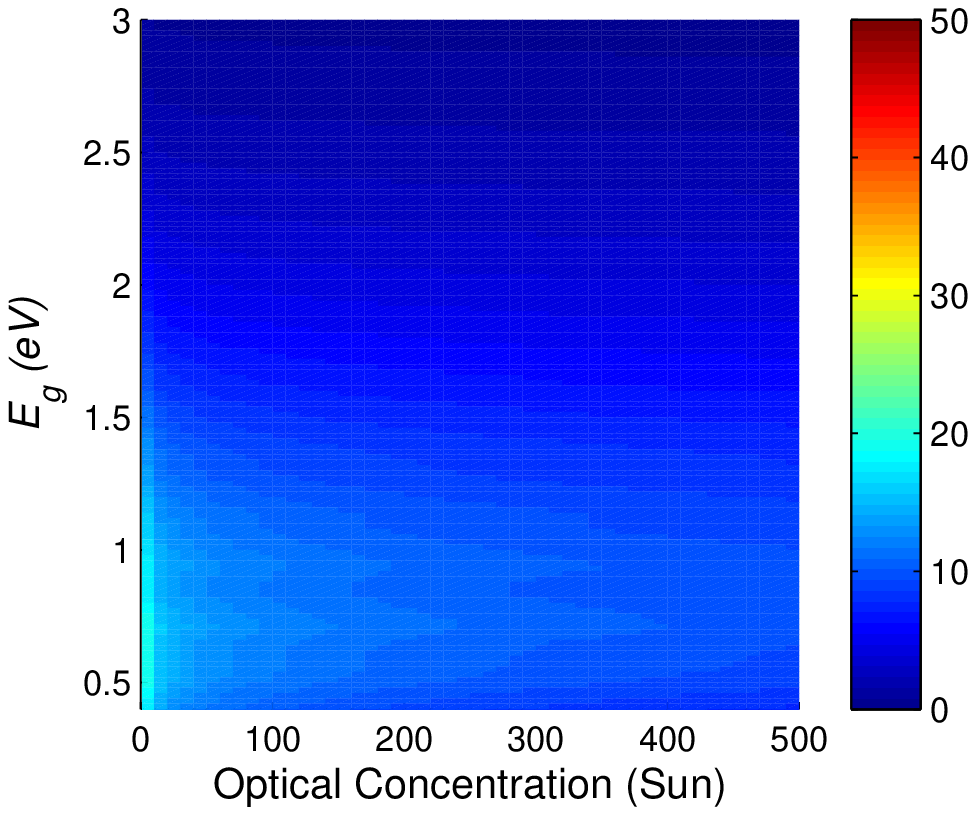}
\caption{The contour of $L_{r}^{(SJ)}$ vs. $E_g$ and $X$ for a single junction solar cell at 300 K.}
\label{SJSCXLTD}
\end{figure}

The other way to reduce relatively the effects of $J_r$ is to concentrate the incident solar radiation. In this case, both $J_g$ and $P_{in}$ increases linearly with $\phi$. So, by $X$ uniform concentration, the conversion efficiency becomes
\begin{equation}
\label{EffX}
	\eta \left( V \right) = \frac{V \, \left( J_g - \dfrac{J_r}{X} \right)}{P_{in}}
\end{equation}
Figure-\ref{SJSCXEff} shows how the efficiency improve with $X$ as $L_{r}^{(SJ)}$ is reduced with increasing $X$ (Figure-\ref{SJSCXLTD}). At 500 sun concentration, $\eta_{max}$ gets to 40.04\% at $1.12 \; eV$ gap.

\begin{figure} [ht]
\centering
\includegraphics[width=3.0in]{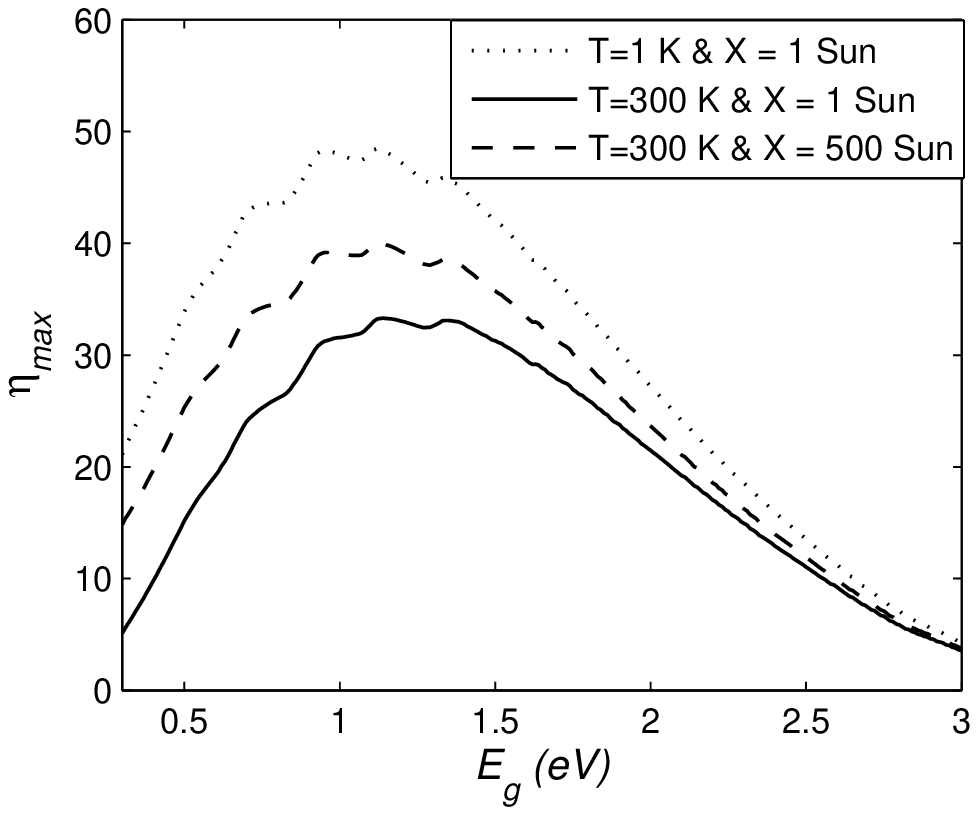}
\caption{$\eta_{max}$ vs. $E_g$ for a single junction solar cell at different optical concentration and temperature.}
\label{SJSCEEff}
\end{figure}

\begin{figure} [ht]
\centering
\includegraphics[width=3.0in]{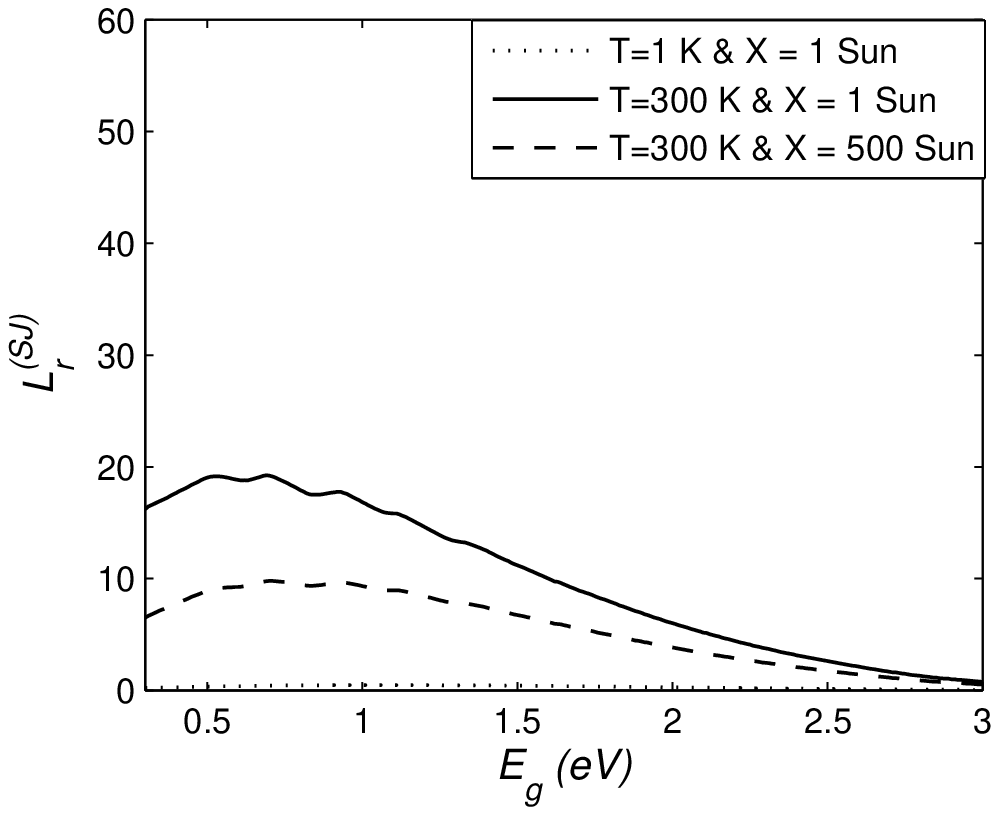}
\caption{$L_{r}^{(SJ)}$ vs. $E_g$ for a single junction solar cell at different optical concentration and temperature.}
\label{SJSCELTD}
\end{figure}

Clearly, the effect of the temperature on $\eta_{max}$ is more than that of the concentration. This is further represented in Figure-\ref{SJSCEEff} and Figure-\ref{SJSCELTD}. At very large $X$, the effects will coincide. Practically, optical concentration is easier to realize. However, it results in more complications as the temperature of the system increases and it commonly results in efficiency reduction and mechanical challenges. This imposes using cooling systems in combined with concentrated solar cells. On the other hand, operating in very low temperature is not practical. However, some device concepts may mimic that.

\subsection{Multi-cell devices}
As shown in the previous section, more than 50\% of energy is lost either due to thermalization or for not absorbed photon. This was known since the early days of solar cells \cite{J03,L01,L03,L12} and many concepts were developed to regain the lost energy. The main concepts were discussion in the Subsection \ref{PVDevConp}. For many concepts like thermalization control and spectrum manipulation ones, there is no general way to estimate the practical efficiency limits. Some models are developed for them; but, they are not practical as they usually result in extreme over-estimation \cite{T03,S11,A05,V04}. For multi-cell devices such as multi-junction and split spectrum, estimating the upper limit can be achieved by extending the single junction model.

Abstractly, in such devices, the spectrum is split and then different cells are used to convert the energy by matching $E_g$. So, there should be $N$ cells with different $E_g^{(i)}$'s (the gaps are ordered ascendantly). Thus, the generated current in the i$^{th}$ cell is
\begin{equation}
\label{Jgn}
	J_g^{(i)} = q \int_{E_g^{(i)}}^{E_g^{(i+1)}} \gamma(E) \phi (E) dE
\end{equation}
For the last cell, the upper limit goes to $\infty$. Similar to the case of single junction devices, the recombination current is calculated based on the generalized black body radiation and it is
\begin{equation}
\label{Jrn}
	J_r^{(i)} = q a \int_{E_g^{(i)}}^\infty  \frac{E^2}{\exp\left( \frac{E - \gamma(E) V^{(i)}}{k T} \right) - 1}  dE
\end{equation}
So, the net generated current in the i$^{th}$ cell is
\begin{equation}
\label{Jn}
	J^{(i)} \left( E_g, V, T \right) = J_g^{(i)} - J_r^{(i)}
\end{equation}

Then for each cell, the conversion efficiency becomes
\begin{equation}
\label{Effn}
	\eta^{(i)} \left( V^{(i)} \right) = \frac{V^{(i)} \, J^{(i)}}{P_{in}}
\end{equation}

In split spectrum cells, the photo-generated current is extracted separately for each cell. So, the total efficiency is 
\begin{equation}
\label{Effss}
	\eta^{(SS)} = \sum_i \eta^{(i)} = \frac{1}{P_{in}} \sum_i V^{(i)} \, J^{(i)}
\end{equation}
For multijunction and intermediate band cells, the photo-generated current should flow from one cell to the other before extracted in the external terminals. This series connection imposes that the current should be the same in all of used cells and it is equal to the lowest current achieved by any of the cells. Thus, the optimization becomes two-step problem. In the first one, $V^{(i)}$ are varied to optimize $\eta^{(i)}$ for each cell. Then the lowest achievable current $J_{min}$ is maintained and hence the $V^{(i)}$ and $\eta^{(i)}$ are changed accordingly. This is done iteratively to maximize the net efficiency which reads
\begin{equation}
\label{Effmj}
	\eta^{(MJ)} = \sum_i \eta^{(i)} = \frac{1}{P_{in}} J_{min} \sum_i V^{(i)}
\end{equation}

\begin{figure} [ht]
\centering
\includegraphics[width=3.0in]{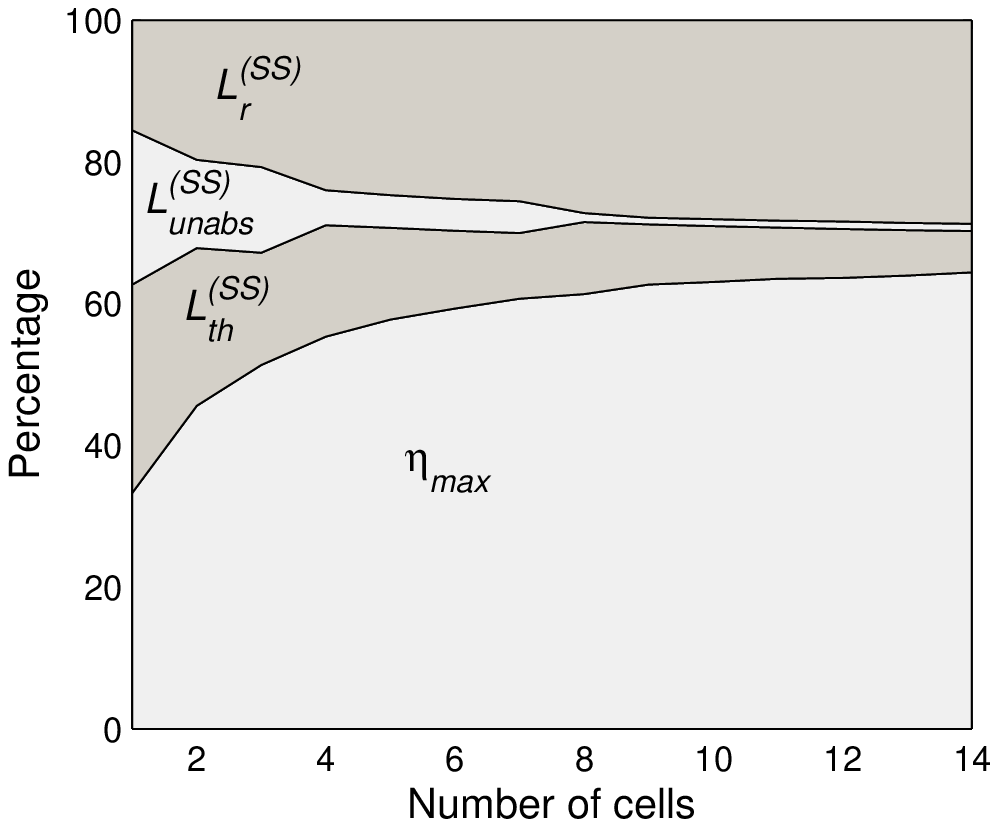}
\caption{The optimized $\eta$ of split spectrum solar cell system for different number of cells at $T = 300 K$ and with no carrier multiplication. The corresponding $L_{unabs}^{(SJ)}$, $L_{th}^{(SJ)}$, and $L_{r}^{(SJ)}$ for each $E_g$ are added up.}
\label{SSSCEffLosses}
\end{figure}

\begin{figure} [ht]
\centering
\includegraphics[width=3.0in]{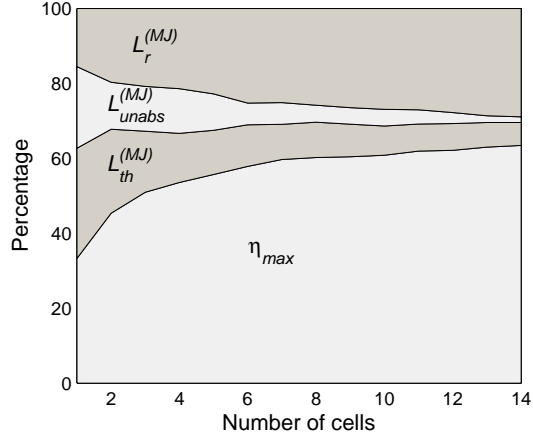}
\caption{The optimized $\eta$ of multi-junction solar cells for different number of cells at $T = 300 K$ and with no carrier multiplication. The corresponding $L_{unabs}^{(SJ)}$, $L_{th}^{(SJ)}$, and $L_{r}^{(SJ)}$ for each $E_g$ are added up.}
\label{MJSCEffLosses}
\end{figure}

\begin{figure} [ht]
\centering
\includegraphics[width=3.0in]{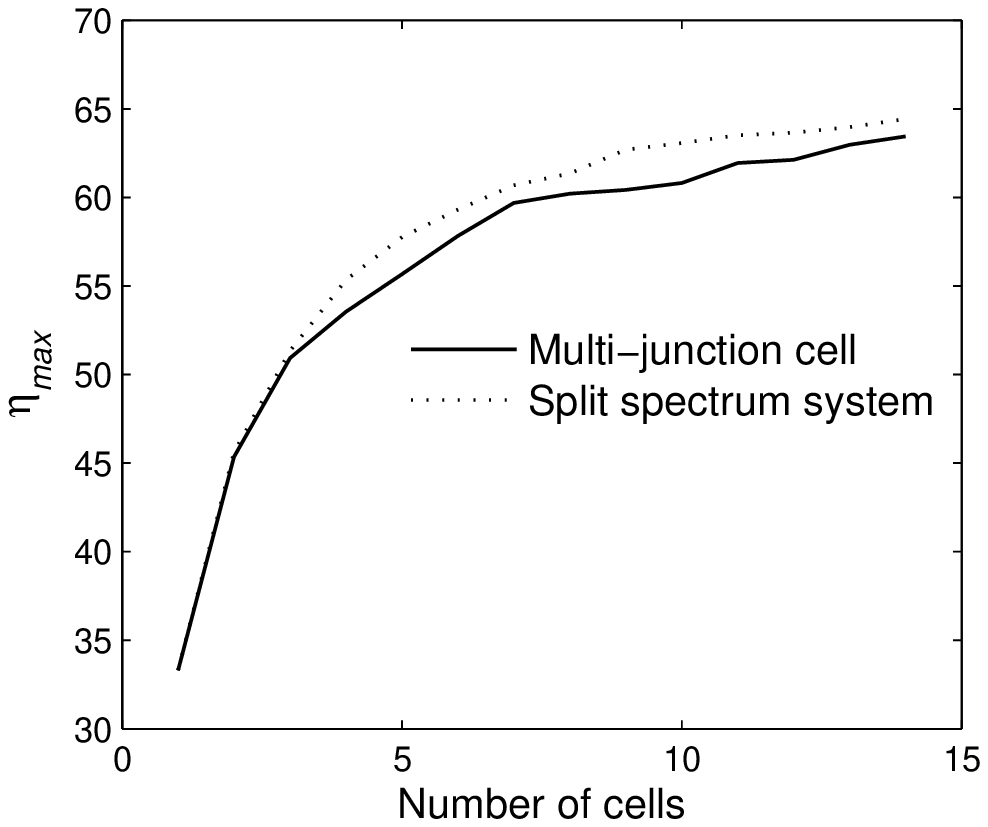}
\caption{$\eta_{max}$ of multi-junction solar cells and split spectrum solar cell system for different number of cells at $T = 300 K$ and with no carrier multiplication.}
\label{CompMJSSEff}
\end{figure}

At the beginning, $\eta$ is optimized for operation at $T = 300 K$ and with no carrier multiplication for different numbers of cells where $E_g$'s are varied for optimization. Figure-\ref{SSSCEffLosses} shows the results for split spectrum cell system and Figure-\ref{MJSCEffLosses} shows the results for multi junction cell, where the corresponding $L_{unabs}$, $L_{th}$, and $L_{r}$ are calculated as well. As expected, $\eta$ increases in both cases, but it is slightly better for split spectrum system as the constraint of current continuity is not a problem. This is shown clearly in Figure-\ref{CompMJSSEff}. The difference in $\eta_{max}$ between the two concept is larger when the number of cells is between 4 and 12. The difference is then diminished for the unpractical larger number of cells. For very large number of cells, the theoretical $\eta$ limit approaches 66\%.

The concept of multi-cell devices were introduced to reduce $L_{unabs}$ and $L_{th}$, which certainly get less with the number of cells. On the other hand, $L_{r}$ increases considerably. This is because the gained energy by reducing $L_{unabs}$ and $L_{th}$ effects are distributed between the extracted photo-generated and radiatively recombined carriers.  

\subsection{Thermalization control based devices}
For thermalization control based devices, there is no general way to estimate efficiency limits as different concepts are applied to control (or avoid) thermalization. In the following, we will discuss the devices based on CM. To raise solar cell efficiencies beyond Shockley-Queisser limit , CM route is proposed \cite{K09,K10,W08,W09,D06,N03}. As aforementioned, the concept of CM is that an energetic photon is utilized to generate multiple electron-hole pairs before it relaxes. Theoretically, many mechanisms can results in CM such as impact ionization \cite{F03,L14}, coherent superposition of multiexcitons \cite{M11,S12}, singlet fission \cite{J06,J07}, and multiexciton generation through virtual states \cite{S13}. Actually, many of these mechanisms are highly correlated.

Experimentally, many groups demonstrate CM for both bulk and nanoscale semiconductor systems. The highest obtained multiplication is seven times in lead salt quantum dots (QDs) by Schaller and Klimov \cite{S08}. However, the experimental apparatuses in most CM experiments are very sophisticated and do not resemble practical PV operations. It is very common in such experiments to induce sequential absorption and force impact ionization by high excitation and strong bias voltage.
The significance of CM and its applicability in solar cells are questioned by many \cite{T02,FH03}. Practically, results are irreproducible in many cases \cite{P05,B06}. For example, Pijpers et al. \cite{P06} rebutted their own earlier results \cite{P07} that overvalued the measured CM. Also, to have significant enhancement in solar cells, CM should be almost ideal. Obviously, this is not the case and it is always noticed that there is an energy threshold preceding the multiplication and that the multiplication is not perfect. It has been shown that these facts limit almost completely any possible enhancement \cite{T02,FH03}. 

Conceptually, the ideal CM condition is when
\begin{equation}
\label{QE01}
	\gamma(E) = \left[ \frac{E}{E_g} \right]
\end{equation}
where the square brackets represent rounding to the lower integer. By considering the fact that energy threshold ($E_{th}$) is always needed before CM starts. Then, the multiplication $\gamma(E)$ becomes
\begin{equation}
\label{QE02}
	\gamma(E) = \theta \left( E - E_g \right) + \theta \left( E - E_{th} \right) \left[ \frac{E - E_{th} + E_g }{E_g} \right]
\end{equation}
where $\theta$ is the Heaviside step function. Furthermore, CM is not perfect and the increase of $\gamma$ over $E/E_g$ is less than one. So, the multiplication would be
\begin{equation}
\label{QE03}
	\gamma(E) = \theta \left( E - E_g \right) + \theta \left( E - E_{th} \right) \left( \frac{E - E_{th} }{E_g} \right) \lambda
\end{equation}

Many models have been used to estimate both $E_{th}$ and $\lambda$. Alharbi \cite{FH03} suggested using empirical relation extracted from experimental data to estimate $E_{th}$. This relation will be used in this review and it is
\begin{equation}
\label{LinApp}
	E_{th} = E_{th,0} + (1+f) \, E_g
\end{equation}
where both $E_{th,0}$ and $f$ are positive. They are interpolated from experimental data to fit the measured $E_{th}$ for different materials. Over solar radiation spectrum, the fitting should ensure that $E_{th} \geq 2 E_g$. For lead salt, it is found that $E_{th,0} = 1.2565$ and $ f = 0.3604$  for PbS and $E_{th,0} = 1.3493$ and $ f = 0.4005$ for PbSe \cite{FH03}.

\begin{figure} [ht]
\centering
\includegraphics[width=3.0in]{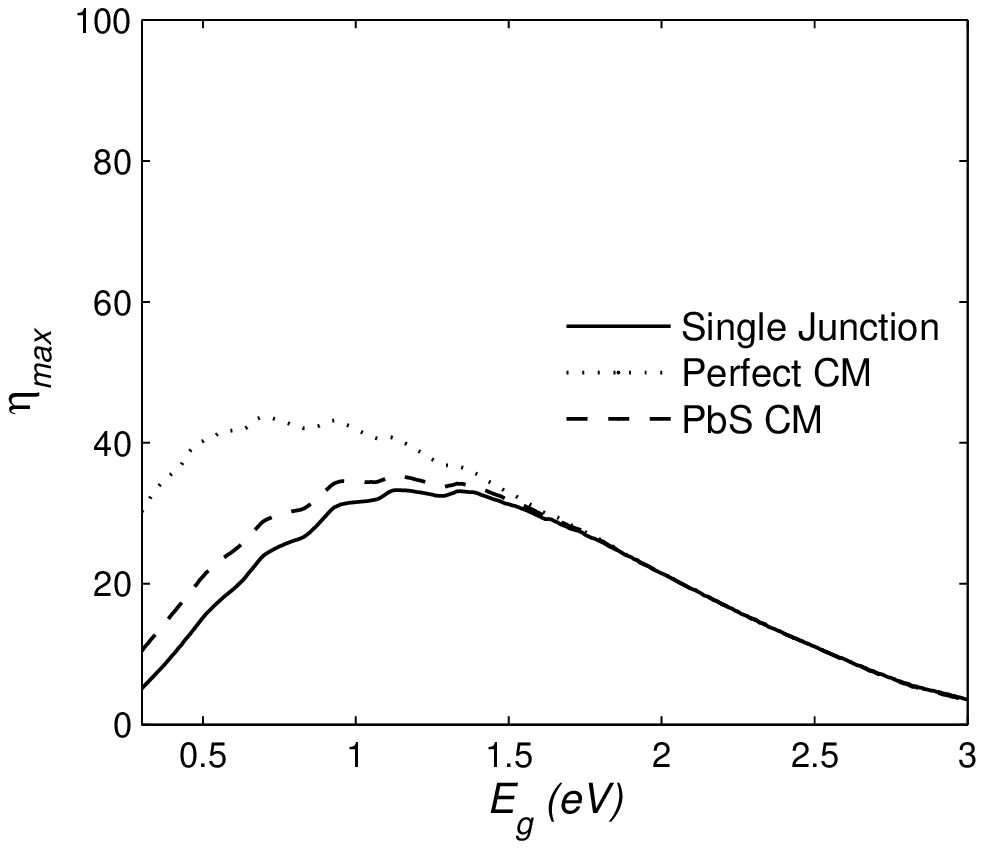}
\caption{The optimized $\eta$ vs. $E_g$ for different CM conditions; namely, No CM, Perfect CM, and the condition for PbS. (to correct the legends)}
\label{CM01Eff}
\end{figure}

\begin{figure} [ht]
\centering
\includegraphics[width=3.0in]{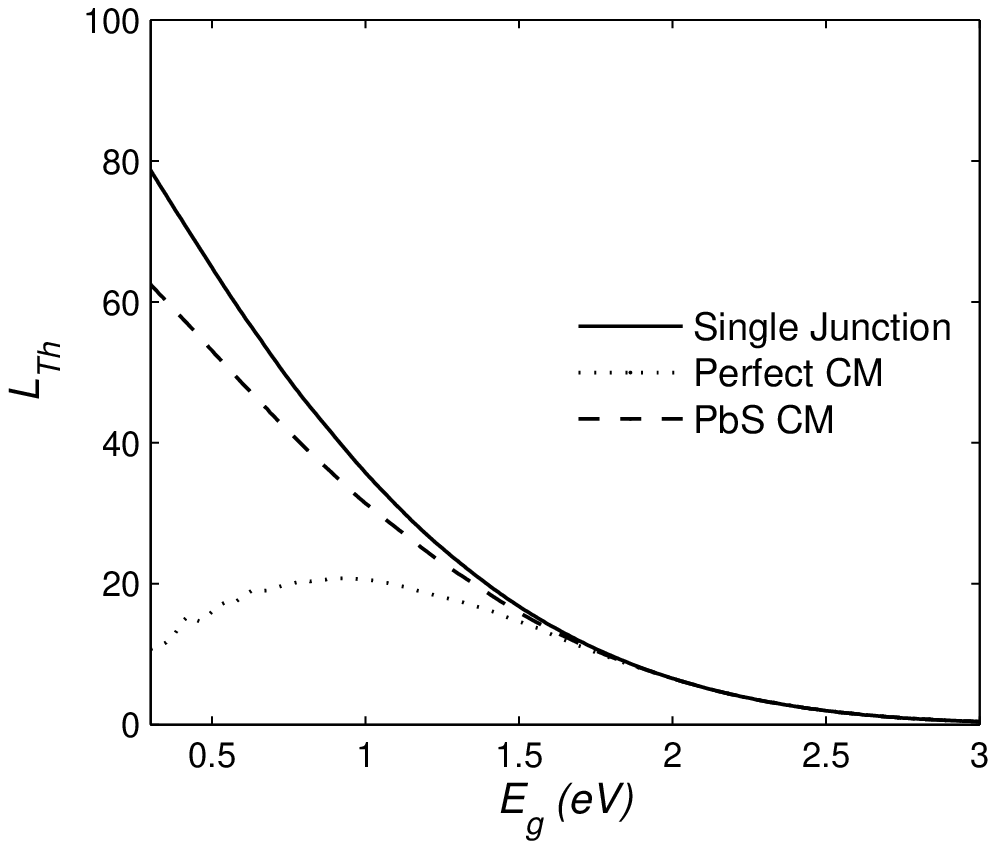}
\caption{The corresponding $L_{th}$ for the optimized $\eta$ vs. $E_g$ for different CM conditions; namely, No CM, Perfect CM, and the condition for PbS. (to correct the legends)}
\label{CM01Lth}
\end{figure}

Figure-\ref{CM01Eff} and Figure-\ref{CM01Lth} show the optimized $\eta$ and the corresponding $L_{th}$ vs. $E_g$ for different CM conditions; namely, No CM, Perfect CM, and the condition for PbS. The condition of PbS is used as its data is the best among the least controversial in term of CM \cite{N04}. It can be observed that more improvement is obtainable for small $E_g$. This is mainly due to the fact that solar radiation start decaying at $3.0 \; eV$ and becomes negligible above $4.0 \; eV$. So, CM becomes negligible for cells with $E_g > 2.0 \; eV$. Theoretically, the efficiency limit can be raised from 33.3\% at $E_g = 1.14 \; eV$ to 44.1\% at $E_g = 0.71 \; eV$ if CM is perfect. However and as aforementioned, this is practically very challenging and it is found that CM advantage for solar cells is severely limited by the needed energy threshold and the imperfect multiplication. To illustrate that, the condition of PbS is used. As can be seen from Figure-\ref{CM01Eff}, the theoretical possible improvement is very marginal.

\begin{figure} [ht]
\centering
\includegraphics[width=3.0in]{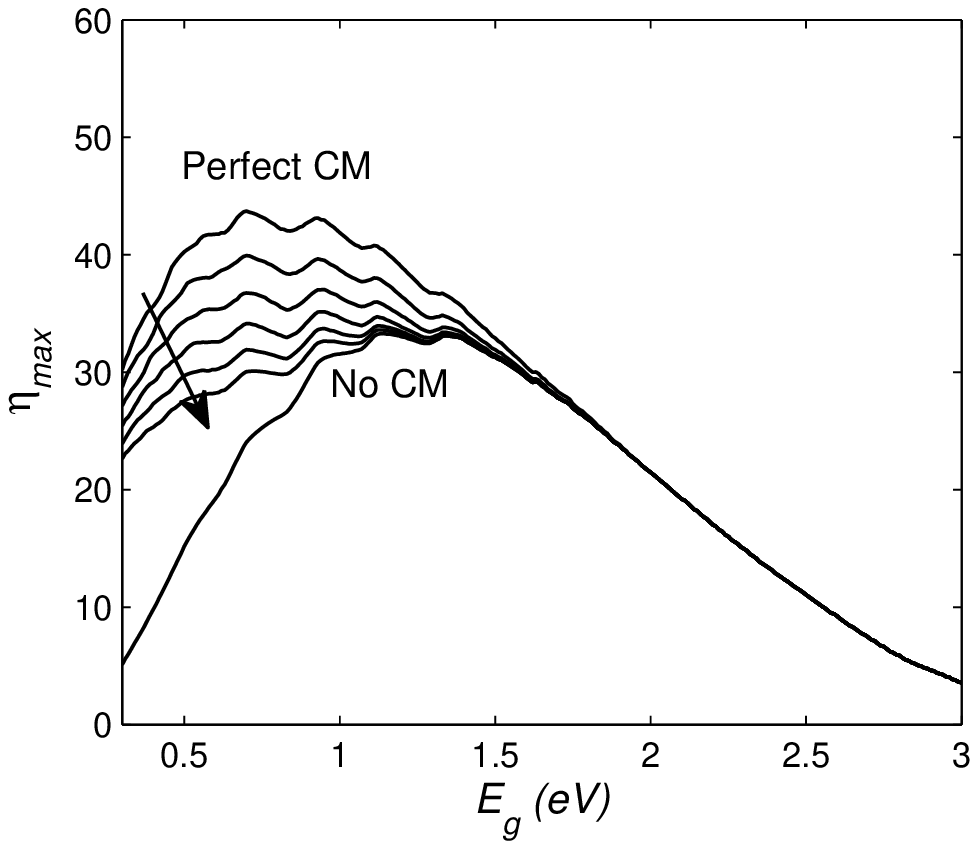}
\caption{The optimized $\eta$ vs. $E_g$ when $E_{th,0}=E_g$ and $f$ is varied from 0 (perfect CM) to 2 at steps on 0.4 along the direction of the arrow.}
\label{CM02Eff}
\end{figure}

\begin{figure} [ht]
\centering
\includegraphics[width=3.0in]{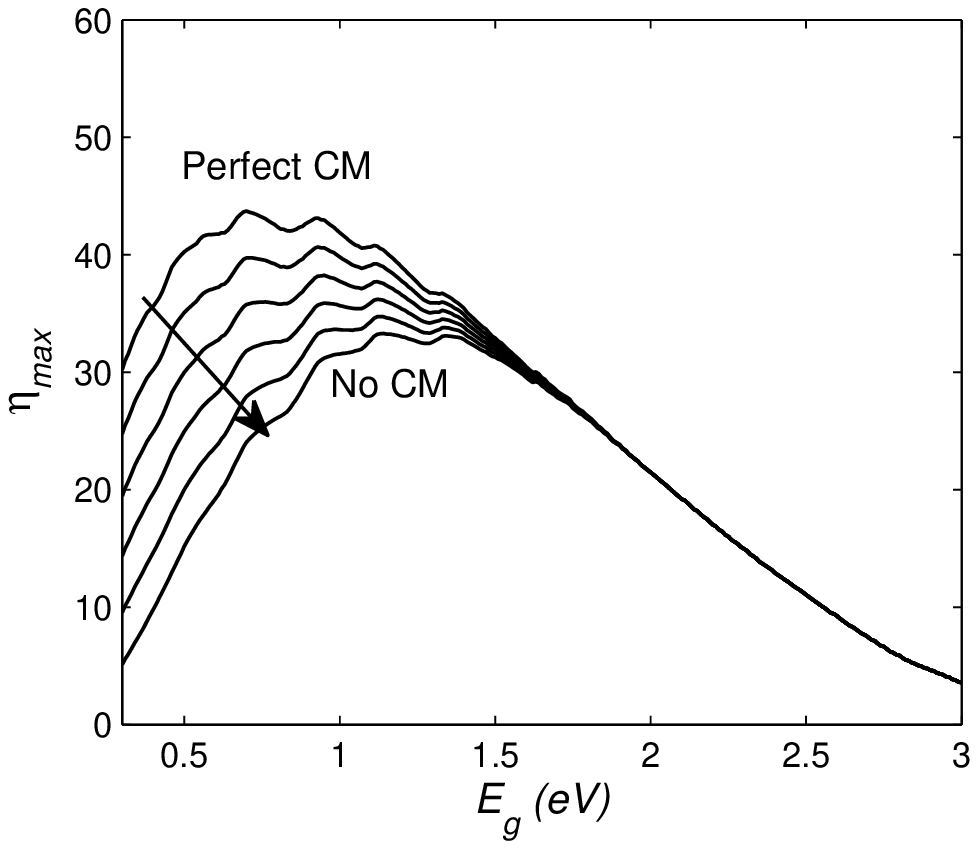}
\caption{The optimized $\eta$ vs. $E_g$ when $E_{th}$ is assumed perfect and equal to $2 E_g$ and $\lambda$ is varied 1 (perfect CM) to 0.2 at steps on 0.2 along the direction of the arrow.}
\label{CM03Eff}
\end{figure}

To illustrate the effects of the needed energy threshold and the imperfect multiplication, $E_{th}$ and $\lambda$ are changed slightly from the perfect condition. In Figure-\ref{CM02Eff}, $E_{th,0}$ is changed by varying $f$ from 0 (perfect CM) to 2 at steps on 0.4 where $E_{th,0}$ is set to 1. It is clear that the advantage of CM is dying rapidly. This is even severer if the multiplication is assumed imperfect as shown in Figure-\ref{CM03Eff}. In this figure, $E_{th}$ is assumed perfect and equal to $2 E_g$. $\lambda$ is varied 1 (perfect CM) to 0.2 at steps on 0.2. $\eta$ dyes very fast with decreasing $\lambda$.

\section{Photosynthesis: Solar-to-chemical energy conversion}
Sunlight is the most abundant energy source available on earth, and therefore to design systems that can effectively gather, transfer or store solar energy has been a great continuing challenge for researchers. To achieve this, a very intuitive approach is to learn from Mother Nature. In light harvesting organisms, the major mechanism that converts light energy into chemical energy is photosynthesis. Photosynthesis can be described by the simplified chemical reaction to obtain carbohydrates:
\begin{equation}
  \mathrm{6CO_2+6H_2O}+N h\nu\rightarrow\mathrm{C_6H_{12}O_6+6O_2}\;.\label{phot-syn}
\end{equation}
Estimates of the efficiency of photosynthesis have a long history 
\cite{photo-eff-1,photo-eff-2,photo-eff-3,photo-eff-4,photo-eff-5,photo-eff-6,photo-eff-7,photo-eff-8a,photo-eff-8b,photo-eff-8c} and depends on how light energy is defined. Ross and Hsiao \cite{Ross-Hsiao} reported that the efficiency cannot exceed $29 \%$ based on  an ideal theoretical  analysis, where entropy and unavoidable irreversibility place a limit on the efficiency of photochemical solar energy conversion. However, photosynthesis is known to occur at $\lambda \leq 700 ~ nm$, where only about $45\%$ of the sun light is photo-synthetically active. Based on these facts, Bolton and Hall \cite{Bolton-Hall} calculated the theoretical maximum efficiency of conversion of light to stored chemical energy in green-plant type photosynthesis in bright sunlight  to be $13.0 \%$ when the principal stable product of photosynthesis is d-glucose. Thermodynamic arguments used in the analysis  which indicate that a photosynthetic system with one photosystem would be highly unlikely to be able to drive each electron from water to evolve   O$_2$  and reduce CO$_2$. The practical maximum efficiency of photosynthesis under optimum  conditions is estimated to be negligible, about few percents $1-3 \%$.  

Remarkably, in plants, bacteria and algae, the photon-to-charge conversion efficiency is about $100\%$ under certain conditions\cite{Book1}. This fact of great interest and generate a lot of excitement to understand how nature optimized the different molecular processes such as trapping, radiative, and non radiative losses, in particular the role of quantum coherence to enhance transport in photosynthesis. This might lead to engineering new materials mimicking photosynthesis and could be used to achieve similar performances in artificial solar cells\cite{Book2,Creatore}. Before discussing in details  the energy transfer in  photosynthesis, first we introduce briefly the concept of quantum coherence which has ignited interest in the possible biological function after the experimental observation of coherence oscillations during energy transfer \cite{LS0}.

\subsection{Quantum Coherence}
In photosynthetic light-harvesting complexes, the electronic coupling between chromophores is similar in magnitude to coupling to the environment and to the disorder in site energies. Thus, quantum effects might influence the dynamics  in these systems.  It is common to examine dynamics of energy transfer in terms of the population evolution from one chromophore to another or from one site to another. Quantum coherence introduces correlations among wave function amplitudes at different sites. The full dynamics should include both the population evolution and coherence accounting for quantum superpositions. In the density matrix formulation, populations described by the diagonal elements and coherence by the off-diagonal elements. 

A quantum state described by a density matrix $\rho$ is called pure if it can be represented by a wave function $\Psi$, $\rho=|\Psi><\Psi|$,  and mixed otherwise. The off-diagonal elements of the density matrix $\rho$ are usually called coherences but they are basis-dependent. To illustrate the concept, let us focus on a system of two excitons  described by  $\Psi(t)=a \Phi_a + b \Phi_b$. The time evolution of the density matrix, $\rho(t) =|\Psi(t)><\Psi(t)|$, is given by:
\begin{equation}
\begin{split}
	\rho(t) = & |a|^2  |\Phi_a><\Phi_a| + |b|^2  |\Phi_b><\Phi_b| \\
	& +a b^* e^{-i(E_a-E_b) t/\hbar } 
|\Phi_a><\Phi_b| \\ 
&+ a^* b   e^{+i(E_a-E_b) t/\hbar }    |\Phi_b><\Phi_a|.
\end{split}
\end{equation}

The first two diagonal  terms represent populations in the excitonic basis whereas the latter two off-diagonal describe coherences.  The phase factors in the off-diagonal elements are responsible for quantum beating.  The frequency of this beating corresponds to the energy difference between the two excitons giving information about the coherence between different chromophores. Recently, Kassal et. al. \cite{Kassal}, in order to address the question if coherence enhance  transport in photosynthesis, they introduce the distinction between state coherence and process coherence. They argue that although some photosynthetic pathways are partially coherent processes, photosynthesis in nature proceeds through stationary states\cite{Kassal}.

\subsection{Photosynthesis}
In photosynthesis, the sunlight is absorbed and excites the electronic states of pigments in the antenna complexes. These electronic excitations then propagate to the reaction center and induce an electron transfer to the primary electron-acceptor molecular called pheophytin. This light to charge conversion is highly efficient and thus illustrates the importance of understanding this excitation energy transfer process in light-harvesting complexes (LHC).

Many light harvesting microbes such as green sulfur bacteria and purple bacteria have been studied as model organisms of photosynthesis. For example, in green sulfur bacteria the most commonly studied LHC is the Fenna-Matthews-Olson (FMO) complex. The FMO complex is situated between the  antenna and the reaction center and functions as an energy pipeline between the two. If the excitonic energy transfer in such LHC can be understood thoroughly, it will be possible to design an artificial light-harvesting system with high efficiency based on a similar mechanism.

Recent experimental results show that long-lived quantum coherence are present in various photosynthetic complexes\cite{K128,K129,K130}. One such protein complex, the Fenna-Matthews-Olson(FMO) complex from green sulphur bacteria\cite{K131,K132}, has attracted considerable experimental and theoretical attention due to its intermediate role in energy transport. The FMO complex plays the role of a molecular wire, transferring the excitation energy from the light-harvesting complex (LHC) to the reaction center(RC) \cite{K133,K134,K135}. Long-lasting quantum beating over a time scale of hundreds of femtoseconds has been observed \cite{LS0,K137}. The theoretical framework for modelling this phenomena has also been explored intensively by many authors \cite{K138,K139,K140,K141,K142,
K143,K144,K145,K146,K147,K148,K149,K150,K151,K152,K153,K154,K155,K156,K157,K158,K159,K160,K161,K162,K163}. 

The fundamental physical mechanisms of energy transfer in photosynthetic complexes is not yet fully understood. In particular, the role of  surrounding photonic and phononic environment on the efficiency or sensitivity of these systems for energy transfer. One major problem in studying light-harvesting complexes has been the lack of an efficient method for simulation of their dynamics under realistic conditions, in biological environments. There are mainly three methods to study the dynamics of such complex open quantum system: (1) The semicalssical Forster method in which the electronic Coulomb interaction among the different chromophores is treated perturbatively (2) The Redfield or Lindblad method in which the electron-phonon interaction is treated perturbatively and (3) The Hierarchy equation of motion method for the intermediate regime when the strength of the Coulomb and electron-phonon interactions are comparable \cite{K153}. The full dynamics can not be treated in such complex systems, thus people relay on simple model Hamiltonians interacting with an approximate environment. The total system Hamiltonian can simplify and written as 
\begin{equation}
H_{total}= H_S+H_B+H_{SB}
\label{FullHamilt}
\end{equation}
where the Hamiltonian for the system 
\begin{equation}
\mathcal{H}_{S}=\sum_{j=1}^{N}\varepsilon_{j}\,\ket{j}\bra{j}+\sum_{j\neq k}J_{jk}\left(\ket{j}\bra{k}+\ket{k}\bra{j}\right), \label{eq:HS}
\end{equation}
where $\varepsilon_{j}$ represents the  excitation energy of the $j$th chromophore (site) and $J_{jk}$ denotes the excitonic coupling between sites $j$ and $k$. The non-nearest neighbour coupling between $j$ and $k$ is treated by the dipole-dipole interaction. 

The environment is described as a phonon bath, modelled by an infinite set of harmonic oscillators:
\begin{equation}
\mathcal{H}_{B} =\sum_{j=1}^{N}\mathcal{H}_{B}^{j}=\sum_{j=1}^{N}\sum_{\xi=1}^{N_{jB}}\frac{P_{j\xi}^{2}}{2m_{j\xi}}+\frac{1}{2}m_{j\xi}\omega_{j\xi}^{2}x_{j\xi}^{2}\;, \label{eq:HB}
\end{equation}
where $m_{j\xi}$, $\omega_{j\xi}$, $P_{j\xi}$, $x_{j\xi}$ are mass, frequency, momentum and position operator of the $\xi$-th harmonic bath associate with the $j$-th site respectively.

The Hamiltonian of the environment $\left( \mathcal{H}_{B} \right)$ and system-environment coupling $\left( \mathcal{H}_{SB} \right)$ can be written as
\begin{equation}
\mathcal{H}_{SB} =\sum_{j=1}^{N}\sum_{\xi}c_{j\xi}\ket{j}\bra{j}x_{j\xi} = \sum_{j=1}^{N}\mathcal{V}_{j}F_{j}, 
\label{eq:HSB}\\
\end{equation}
where $\mathcal{V}_{j}=\ket{j}\bra{j}$ and $F_{j}=\sum_{\xi}c_{j\xi} x_{j\xi}$. Here,  $c_{j\xi}$ represents the system-bath coupling constant between the $j$-th site  and $\xi$-th phonon mode, here we assume each site  is coupled to the environment independently. The dynamics of such an open quantum system is given by the quantum Liovillian equation and one has to relay on approximations depends on the different coupling terms.

\begin{figure*} [ht]
\centering
\includegraphics[width=5.0in]{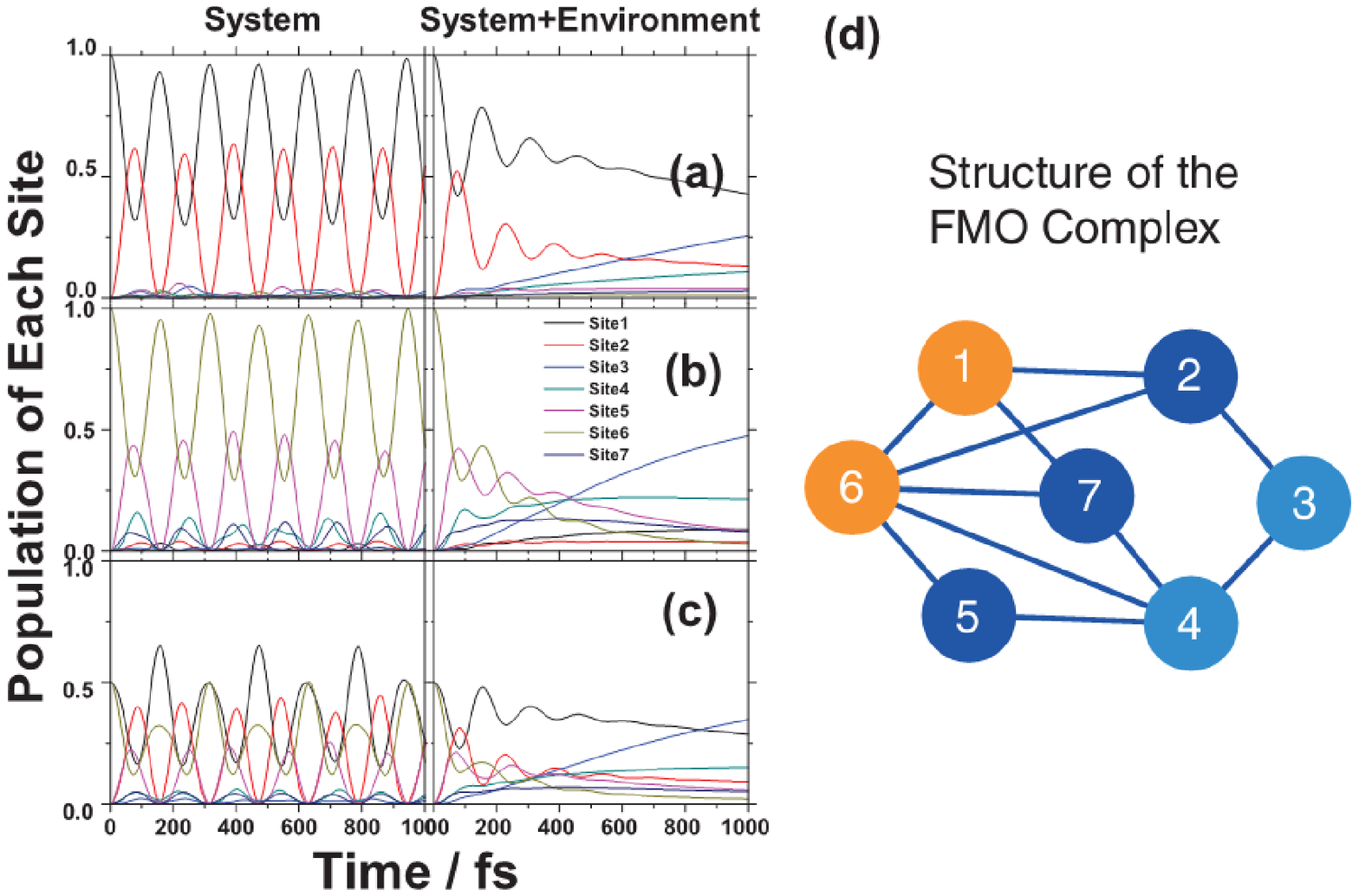}
\caption{The quantum  evolution for the site population  in the FMO
complex  of each site at cryogenic temperature  T = 77 K. The left panel
shows the dynamics for the system alone and the right includes the effects of the environment.
The reorganization energy is $\lambda_j=\lambda= 35 cm^{−1}$, while the value of Drude decay constant is
$\gamma_j^{-1}=\gamma^{-1}= 50 fs$. The initial conditions are site 1 excited (a), Site 6 excited (b) and the
superposition of site 1 and  6 (c). }
\label{FMO}
\end{figure*}

To examine quantum coherence, Jing et. al. \cite{K143}  used a  new developed modified scaled hierarchical approach, based on the model Hamiltonian defined above in Eq. \ref{FullHamilt}, and show that the time scales of the coherent beating are consistent with experimental observations \cite{LS0}. Furthermore, the theoretical results exhibit population beating at physiological temperature. Additionally, the method does not require a low-temperature correction to obtain the correct thermal
equilibrium at long times. The results for the FMO complex are presented in Figure-\ref{FMO}. On the left panel, we show the results of simulation for the system Hamiltonian only. The the right panel one observes that the quantum beating between certain sites clearly persists in the short time dynamics of the full FMO complex \cite{K143,K162}. For the simulated initial conditions, the population beatings can last for hundreds of femtoseconds; this time scale is in agreement with the experimental observation \cite{LS0}.

\begin{figure*} [ht]
\centering
\includegraphics[width=6.0in]{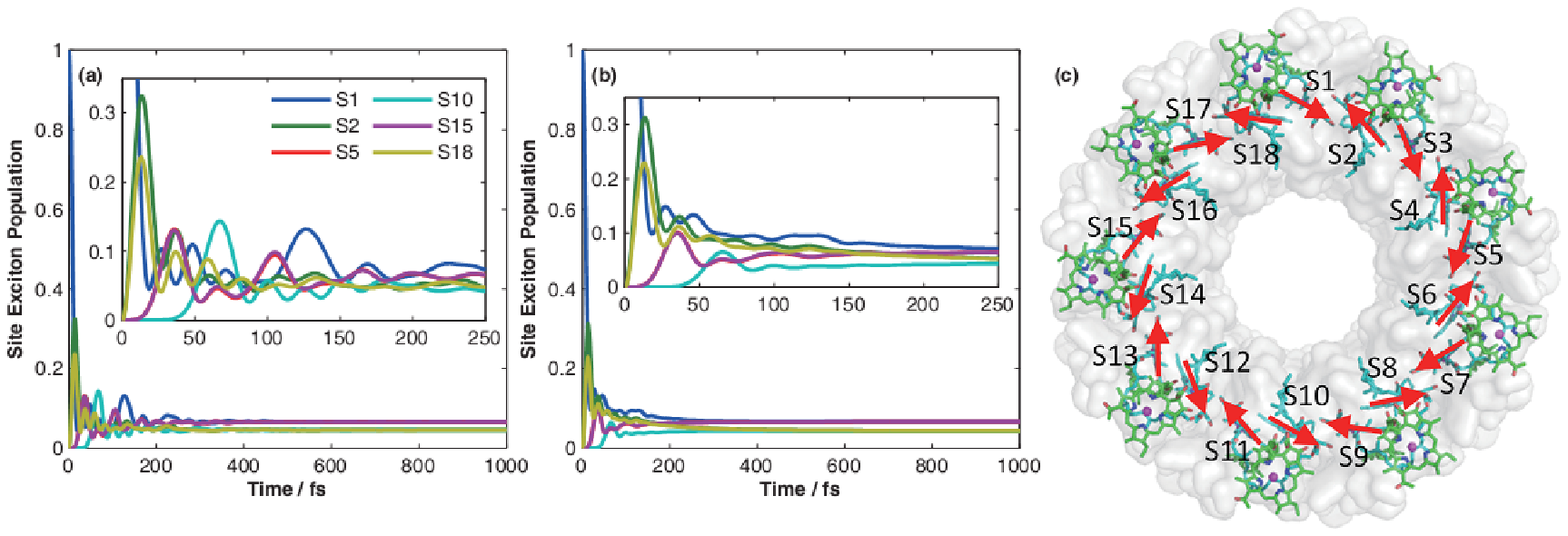}
\caption{The excitation population dynamics of LH2 B850 18 sites at 77 K.
The exciton population dynamics of B850 bacteriochlorophylls (BChls)
with site 1 (S1) initially excited. (a) The population evolution of S1, S2, S5, S10, S15, and
S18 without dissipation (the system is isolated and uncoupled to bath). (b) The population
dynamics of the same sites while the system is coupled to bath at room temperature T =
300 K. The coherent energy transfer lasts about 150 fs and the whole system is equilibrated
after 400 fs. The inset is a magnification of the first 250 fs dynamics.}
\label{LH2}
\end{figure*}

Recently, the modified scaled hierarchical approach was used also by Shuhao et. al \cite{K161} to examine the electronic excitation population and coherence dynamics in the chromophores of the
photosynthetic light harvesting complex (LH2) B850 ring from purple bacteria (Rhodopseudomonas
acidophila). The  oscillations of the excitation population and coherence in the site basis are also observed in LH2. However, this oscillation time (300 fs) is much shorter compared to the FMO protein (650 fs) at cryogenic temperature. Both environment and high temperature are found to enhance the propagation speed of the exciton wave packet yet as expected they shorten the coherence time and suppress the oscillation amplitude of coherence and the population. In Figure-\ref{LH2} we show the numerical results of the  excitation population dynamics for LH2 B850 18 sites at 77 K.

Here we are dealing with  the exciton transport, excitons are quasiparticles, each formed from a pair of electron and hole, that provide a natural means to convert energy between photons and electrons. There are several possible ways to measure the success rate of an energy transfer process \cite{K165}, such as energy transfer efficiency and transfer time. In order to examine the 
transport efficiency one should calculate the exciton recombination and exciton trapping. These can be calculated by \cite{K165}:
\begin{equation}
H_{trap}=-i \hbar \sum_j       \kappa_j   \ket{j}\bra{j}; \;\;  H_{recomb}=-i \hbar \Gamma \sum_j  \ket{j}\bra{j}
\end{equation}
Where $\Gamma$ is the rate of recombination at every site and is trapped with a rate $\kappa_j$ at certain molecules. The probability that the exciton is captured at a certain  $j$th site within the time interval $(t,t+dt)$ is given by  $2 \kappa_j \bra{j} \rho(t) \ket{j} dt$. Thus, the efficiency can be defined as
\begin{equation}
\eta=  2 \sum_j \kappa_j \int_0^{\infty} \bra{j} \rho(t) \ket{j} dt
\end{equation}
which is the integrated probability at different sites. The other  measure for the quantum transport is the average transfer time which is defined as
\begin{equation}
\tau=  \frac{2}{\eta}  \sum_j \kappa_j \int_0^{\infty} \bra{j} \rho(t) \ket{j} t dt
\end{equation}

Recently, Rebentrost et al. \cite{K165} have argued that at low temperatures, the dynamic is dominated by coherent hopping, the system is disordered and exhibit quantum localization, depending on the variation in the site energies. Once forming an excitonic state localized at an initial site, coherent hopping alone has a low efficiency in transporting the excitation from the initial site to another site with significantly different energy. However, interaction with the environment leading to dephasing can destroy the excitation localization and enhance transport. Thus, decoherence might enhance transport if the dephasing rate does not grow larger than the terms of the system Hamiltonian.  The idea that decoherence enhance transport used to explain the high efficiency of  excitonic transport using  FMO protein of the green sulfur bacterium \cite{K165}.  

\subsubsection{Limits of quantum speedup in photosynthesis}
After discovering experimentally the long lived quantum coherence in photosynthetic light-harvesting complexes, it was suggested that excitonic transport features speedups analogous to those found in quantum algorithms. In particular a Grover quantum  search type speedup \cite{LS0}. Whaley et al. \cite{LS1}  investigated  this suggestion by comparing the dynamics in these systems to the dynamics of quantum walks. They have found that the speedup happens at very short time scale (70 fs) compared with the longer-lived quantum coherence (ps scale). To distinguish between quantum speedup and classical diffusive transport one can calculate the exponent, $n$, of the power law for the mean-squared displacement $<x^2>$  as a function of time, $t$. The mean-squared displacement can be obtained from the density matrix $\rho$ of the system,  $<x^2>=\mathrm{Tr} (\rho x^2)/ \mathrm{Tr} (\rho)$. To obtain the 
exponent $n$ in $ <x^2> \approx t^n$, one examine the slope of the log-log plot of the mean-squared disparagement vs. time. If the exponent $n=1$, this corresponds to the limit of diffusive transport, whereas and exponent $n=2$ corresponds to ideal quantum speedup, a ballistic transport. Using the FMO complex, which acts as a quantum wire, with a seven site Hamiltonian with the parameters calculated by Adolphs and Renger \cite{LS2}, Whaley et. al. \cite{LS1} have found that best fit for the exponent was $n=2$ happens at short time (about 70 fs), then a transition from ballistic to sub-diffusive transport with $n=1$ though quantum coherence lasts over 500 fs in their model of calculations. The short lived nature of quantum speedup ( about 70 fs)  might implies that the natural process of energy transfer in these photosynthetic complexes does not correspond to quantum search. Their results suggest that quantum coherence effects in photosynthetic complexes are optimized for high efficiency of transporting the excitation from the antenna to the reaction center and not for the goal of quantum speedup.  

\subsubsection{Excitonic diffusion length in complex quantum systems}
It is well known that the phenomenon of superradiance, introduced by Dike \cite{ED1}, formed by a quantum interference effect induced by symmetry. Due to this cooperative phenomena, the probability of a single photon emission from N identical atoms collectively interacting with vacuum fluctuations becomes N times larger than incoherent individual spontaneous emission probabilities \cite{ED2} and studied in detail in \cite{ED3,ED4,ED5,ED6}. The same basic mechanism could lead to an analogous phenomenon known as cooperative energy transfer or supertransfer \cite{ED7,ED8,ED9}. In particular, the exciton transfer rate under such assumptions. With very strong and symmetrized interactions of N  molecules the excitation becomes highly delocalized, leading to a large effective dipole moment associated with the N molecules, leading to supertransfer. Abasto et. al \cite{ED10} have shown that symmetric couplings among aggregates of $N$ chromophores increase the transfer rate of excitons by a factor $N^2$  and demonstrated how supertransfer effects induced by geometrical symmetries can enhance the exciton diffusion length by a factor N along cylindrically symmetric structures, consisting of arrays of rings of chromophores, and along spiral arrays. It will be of great interest to examine this phenomena in novel excitonic devices since a major problem in design and fabrication  is the limited exciton diffusion length that could be of about 10 nm in disordered materials. This limitation, has lead to low efficiency and complicated device structures in organic photovoltaic cells \cite{ED11}. It will be of interest to examine whether one can use quantum-mechanical supertransfer effects to enhance exciton diffusion length in photovoltaic cells \cite{ED11,ED12,ED13}.

\section{Quantum Coherence: Intuitive aspects for solar energy conversion}
Quantum mechanics which was developed in the twentieth century continues to yield new fruit in the twenty-first century. For example, quantum coherence effects such as lasing without inversion \cite{Lasing,Lasing1}, the Photo-Carnot Quantum heat engine \cite{Carnot},Photosynthesis  and the
quantum photocell \cite{Photocell} are topics of current research interest which are yielding new insights  into thermodynamics and optics.

\begin{figure} [ht]
\centering
\includegraphics[width=3.0in]{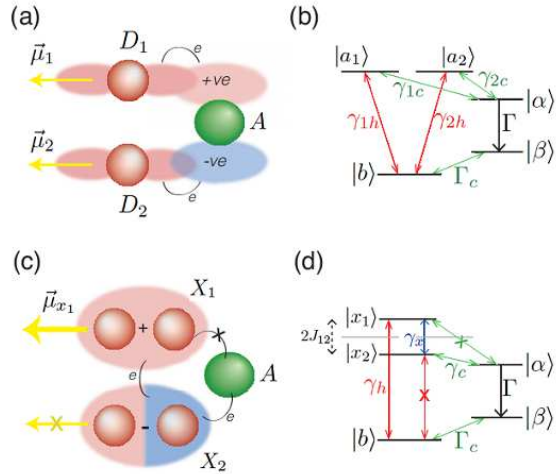}
\caption{The photosynthetic reaction centers used in the scheme proposed by Creatore et al. \cite{Creatore} to enhance photon to current conversion by suppressing the recombination. In (a), the doners $D_1$ and $D_2$ are identical, but uncoupled. In this case, the recombination is not suppressed and the rates $\gamma_{1h}$ and $\gamma_{2h}$ are equal. (b) is the level scheme of this case. In (c), the coupling between the doners results in coupled eigenstates ($X_1$ and $X_2$) due to the symmetric and antisymmetric superposition of the original doner states. The level scheme of this system is shown in (d). The analysis show that recombination is suppressed as a result of the coupling. (Copied with permission from the original paper: PRL 111, 253601 (2013) \cite{Creatore}).}
\label{ChinPRL1}
\end{figure}

\begin{figure} [ht]
\centering
\includegraphics[width=3.0in]{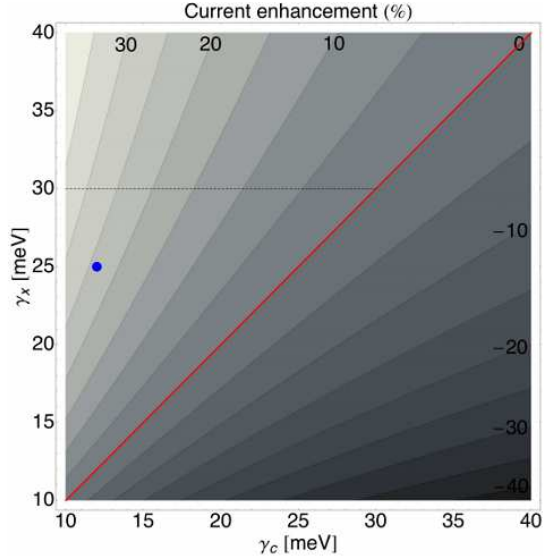}
\caption{The current enhancement due to coupling between the original doners as a function of various relaxation rate ($\gamma_x$) and the electron transfer rate ($\gamma_c$) at 300 $K$. ((Copied with permission from the original paper: PRL 111, 253601 (2013) \cite{Creatore}).}
\label{ChinPRL2}
\end{figure}

Photosynthesis is one of the most common phenomenon in nature, but the detailed principles of the whole process are still unclear. A more recent and still rapidly expanding field of research studies how quantum physics plays  a much more profound role in solar-energy conversion, notably through various interference and coherence effects. The energy transfer from the light harvesting complex  to the reaction center is amazingly high, almost $100\%$.  Does quantum coherence enhance transport in photosynthesis? Artificially reproducing the biological light reactions responsible for this remarkably efficiency represents a new research direction. Recently, Creatore et. al. \cite{Creatore} developed such a scheme  and present a model photocell based on the nanoscale architecture of photosynthetic reaction centres that explicitly harnesses the quantum mechanical effects recently discovered in photosynthetic complexes. They show that Quantum interference
of photon absorption/emission induced by the dipole-dipole interaction between molecular excited states, as shown in in Figure-\ref{ChinPRL1}, guarantees an enhanced light-to-current conversion and hence power generation for a wide range of realistic parameters. The enhancement in the current is shown in Figure-\ref{ChinPRL2}. This shall open a promising new route for designing artificial light-harvesting devices inspired by biological photosynthesis and quantum technologies. They show that the naturally-occurring dipole-dipole interactions between suitably arranged chromophohores can  generate quantum interference effects that can enhance the photo-currents and maximum power outputs by $ > 35 \%$ over a classical cell.

Quantum heat engines convert hot thermal radiation into low-entropy useful work. The ultimate efficiency of such system  is usually governed by a detailed balance between absorption and emission of the hot pump radiation. The laser is an example of such system. Moreover, it was demonstrated both theoretically and experimentally that noise-induced quantum coherence  can break detailed balance and yield lasers without population inversion  with enhanced efficiency \cite{PNAS}. Scully show \cite{Photocell} that it is possible to break detailed balance via quantum coherence which yields a quantum limit to photovoltaic operation which can exceed the classical Shockley-Quiesser recombination limit. The analysis considers a toy photocell model which is constructed to be a counterpart to ``lasing without inversion" . In conventional lasing, one considers an ensemble of two-level atoms (plus additional levels not directly involved in the photoemission process) all coupled to the same cavity mode. These atoms can undergo both absorption as well as the converse emission processes. If, by optical pumping, one achieves a situation in which there are more atoms in the excited than in the ground state (a so-called \emph{inversion}) then emission will dominate over absorption, and one will under certain conditions observe a net emission, i.e., lasing. The idea  is to relax the lasing threshold by suppressing the absorption process. This is achieved by splitting the atoms' ground state into two near-degenerate levels. In quantum mechanics, one cannot simply add the absorption probabilities for these two levels; instead, one has to add the relevant transition amplitudes coherently, and the resulting absorption rate may actually be \emph{less} than the individual rates due to destructive interference. In an elegant corollary, one can now try to suppress the \emph{emission} process in a photovoltaic device (which in this case is the undesirable process, leading to efficiency loss via recombination), by replacing the \emph{upper} level with a near-degenerate doublet, engineering the system parameters such that the two recombination sub-processes interfere destructively. Recently, Dorfman et al. \cite{PNAS}  have introduced a promising approach to this problem, in which the light reactions are analysed as quantum heat engines. Treating the light-to-charge conversion as a continuous Carnot like cycle they  show  that quantum coherence could boost the photo-current of a photocell based on photosynthetic reaction centres by at least $27\%$ compared to an equivalent classical photocell.  

Two seemingly unrelated effects attributed to quantum coherence have been discussed. First, an enhanced solar cell efficiency was predicted and second, population oscillations were measured in photosynthetic antennae excited by sequences of coherent ultrashort laser pulses. Because both systems operate as quantum heat engines that convert the solar photon energy to chemical energy in photosynthesis and to electric current in solar cells. Artificially reproducing the biological light reactions responsible for the remarkably efficient photon-to-charge conversion in photosynthetic complexes represents a new direction for the future development of photovoltaic devices.  


\section{Conclusion}
In this review, we summarized different PV device concepts and their efficiency theoretical limits where more discussion emphasize is toward the losses. It is shown that the efficiency of single-junction PV is at best 33.3\% in normal conditions at 300 $K$. This can be improved by either cooling or optical concentration to 48.48.\% and 40\%+ respectively. However, optical concentration is more practical. Cooling toward very low temperatures is not practical; yet, it can be conceptually mimicked. For multi-cell PV systems, the efficiency can be improved by reducing the losses due to thermalization and  unabsorbed photons. the analysis show that split-spectrum system should result in better efficiency when compared to multijunction and intermediate cells. Though bulky, it is easier to build.

Few lessons from nature and other fields to improve the conversion efficiency in PVs are presented and discussed. From photosynthesis, although it was shown that the whole conversion efficiency of photosynthesis process is not attracted, the perfect exciton transport in photosynthetic complexes can be utilized for PVs. Remarkably, in plants, bacteria and algae, the photon-to-charge conversion efficiency is about $100\%$ under certain conditions. Also, we present some lessons learned from other field that can be used in PVs like recombination suppression by quantum coherence. For example, the coupling in photosynthetic reaction centers is used to suppress recombination in photocells. Theoretical, it can enhanced the net photo-generated current by 35\%.

\section*{Acknowledgement}

We would like to thank Alec Maassen van den Brink for the useful discussions.

\section*{References}

\bibliographystyle{elsarticle-num}
\bibliography{SCEff}

\end{document}